\documentclass[ prx, twocolumn, nofootinbib, floatfix,superscriptaddress]{revtex4-2}

\usepackage{notes2bib}
\usepackage{natbib}
\usepackage{gensymb}
\usepackage{amsthm}
\usepackage{amsmath}
\usepackage{color}
\usepackage{bm}
\usepackage{amssymb}
\usepackage{amsfonts}
\usepackage{dsfont}
\usepackage{graphicx}
\usepackage{float}
\usepackage{subfigure}
\usepackage{verbatim}
\usepackage{amsfonts}
\usepackage{bbold}
\usepackage{dcolumn}
\usepackage{bm}
\usepackage[dvipsnames]{xcolor}
\usepackage{listings}
\definecolor{blueprl}{RGB}{46,48,146}
\usepackage[pdftex,colorlinks=true,urlcolor=blueprl,citecolor=blueprl,linkcolor=blueprl]{hyperref}
\usepackage{times}
\usepackage{color}

\newcommand{\tcb}{\textcolor{blue}}

\usepackage[dvipsnames]{xcolor}
\newcommand{\MB}[1]{\mathbf{#1}}
\newcommand{\h}[1]{\hat{#1}}

\usepackage[mathscr]{euscript}
\usepackage{amsmath}
\usepackage{graphicx}
\usepackage{dcolumn}
\usepackage{bm}
\usepackage{epsfig}
\usepackage{amssymb,latexsym,mathrsfs}
\usepackage{graphicx}
\usepackage{color}
\usepackage{hyperref}
\usepackage{float}
\usepackage{diagbox}

\newcommand{\vt}{\vphantom{\frac{1}{2}}}
\newcommand{\infint}{\int_{-\infty }^{+\infty }}
\newcommand{\mrm}[1]{{\mathrm{#1}}}
\newcommand{\nonvt}{\nonumber \vphantom{\frac{1}{\sqrt{2}}}}

\definecolor{vividviolet}{rgb}{0.62, 0.0, 1.0}
\definecolor{amaranth}{rgb}{0.9, 0.17, 0.31}
\definecolor{palatinateblue}{rgb}{0.15, 0.23, 0.89}
\definecolor{brightpink}{rgb}{1.0, 0.0, 0.5}
\definecolor{cornflowerblue}{rgb}{0.39, 0.58, 0.93}
\definecolor{deepcarminepink}{rgb}{0.94, 0.19, 0.22}
\definecolor{radicalred}{rgb}{1.0, 0.21, 0.37}
\definecolor{blueblue}{RGB}{21,47,181}
\definecolor{greengreen}{RGB}{65,166,16}

\newcommand{\be}{\begin{equation}}
\newcommand{\ee}{\end{equation}}
\newcommand{\bs}{\begin{split}} 
\newcommand{\bea}{\begin{eqnarray}}
\newcommand{\eea}{\end{eqnarray}}
\newcommand{\non}{\nonumber }

\usepackage{mathtools}
\newcommand{\D}{\mathrm{d}}
\newsavebox{\myhbar}
\usepackage[normalem]{ulem}
\newcommand{\dd}{\dagger}
\newcommand{\VT}{\vphantom{\frac{1}{\sqrt{2}}}}

\newcommand{\four}[1]{{\color{violet} #1}}

\begin{document}

\title{Relativity and decoherence of spacetime superpositions}
\author{Joshua Foo}
\email{joshua.foo@uqconnect.edu.au}
\affiliation{Centre for Quantum Computation \& Communication Technology, School of Mathematics \& Physics, The University of Queensland, St.~Lucia, Queensland, 4072, Australia}
\affiliation{Department of Physics, Stevens Institute of Technology, Castle Point Terrace, Hoboken, New Jersey 07030, U.S.A.}
\author{Robert B.\ Mann}
\affiliation{Perimeter Institute, 31 Caroline St., Waterloo, Ontario, N2L 2Y5, Canada}
\affiliation{Department of Physics and Astronomy, University of Waterloo, Waterloo, Ontario, Canada, N2L 3G1}
\author{Magdalena Zych}
\affiliation{Department of Physics, Stockholm University, AlbaNova University Center, SE-106 91 Stockholm, Sweden}
\affiliation{Centre for Engineered Quantum Systems, School of Mathematics and Physics, The University of Queensland, St. Lucia, Queensland, 4072, Australia}

\begin{abstract}
It is univocally anticipated that in a theory of quantum gravity, there exist quantum superpositions of semiclassical states of spacetime geometry. Such states could arise for example, from a source mass in a superposition of spatial configurations. In this paper we introduce a framework for describing such ``quantum superpositions of spacetime states.'' We introduce the notion of the relativity of spacetime superpositions, demonstrating that for states in which the superposed amplitudes differ by a coordinate transformation, it is always possible to re-express the scenario in terms of dynamics on a single, fixed background. Our result unveils an inherent ambiguity in labelling such superpositions as genuinely quantum-gravitational, which has been done extensively in the literature, most notably with reference to recent proposals to test gravitationally-induced entanglement. We apply our framework to the above mentioned scenarios looking at gravitationally-induced entanglement, the problem of decoherence of gravitational sources, and clarify commonly overlooked assumptions. In the context of decoherence of gravitational sources, our result implies that the resulting decoherence is not fundamental, but depends on the existence of external systems that define a relative set of coordinates through which the notion of spatial superposition obtains physical meaning.
\end{abstract} 

\date{\today} 

\maketitle 

\section{Introduction}

The twin discoveries of quantum mechanics and Einstein's theory of general relativity revolutionized our understanding of physical reality. Perhaps the most distinctive features of these respective theories are the notions of quantum superposition and spacetime. The former postulates that a quantum system with a distinct set of possible configurations may also exist in a state featuring many of these configurations ``at once.'' The latter was Einstein's geometric interpretation of gravity as the curvature of a pseudo-Riemannian manifold, generated by the presence of mass and energy \cite{synge1960relativity}. 

The most significant challenge for modern physicists is finding a consistent unification of quantum theory with general relativity. While effective field theory (EFT) approaches \cite{Burgess2004,cornetdoi:10.1142/9789814530347} have been developed and successfully applied to the description of spin-2 gravitons at low energies (and the resulting emergence of classical general relativity in this limit), the problem of how to correctly characterize quantum superpositions of gravitational fields or spacetime geometries remains an open question \cite{dewitt2011role}. It is univocally anticipated that any theory combining relativistic gravitation with quantum mechanics must be able to describe spacetime as possessing quantum-mechanical degrees of freedom, whose states reside in a complex Hilbert space and may be placed in quantum superpositions of different configurations. Various formal frameworks such as the Wheeler-deWitt equation \cite{Wheeler1968SUPERSPACEAT,dewittPhysRev.160.1113} and loop quantum gravity \cite{rovelli2008loop,rovelli2014covariant,thiemann2003lectures} have been proposed, which (in principle) allow  for such environments to be constructed \cite{rovelli2004quantum,KASTRUP1994665,Campiglia_2007,Gambini_2014,gambiniPhysRevLett.110.211301,Kiefer_2013}.

In this article, we focus on the fundamental issue underlying many recent studies, namely the physical meaning of ``superpositions of spacetime metrics'' (similarly superpositions of Newtonian gravitational fields in the nonrelativistic limit). We mainly focus on spatial superposition of a source mass in our examples, but our approach and main results hold for superpositions of states related by any symmetry of dynamics. Spatial superposition states are considered in the proposals for possible observations of gravitationally-mediated entanglement~ \cite{bose2017spin,marletto2017gravitationally}, in the context of decoherence that they may induce on quantum systems \cite{danielson2022black,arrasmith2019,Demers_1996,allaliPhysRevLett.127.031301,Allali_2020,Gambini2007}, and the indefinite causal structures they give rise to \cite{zych2019bell,castroPhysRevX.8.011047,Paunkovic2020causalorders,howl2022https://doi.org/10.48550/arxiv.2203.05861,Brukner2014}. Here we demonstrate that any effects emerging due to a spatial superposition of a massive object can be formally reproduced using a single spacetime metric, due to two basic tenets of quantum mechanics and relativity: linearity of quantum theory and invariance of dynamics under general coordinate transformations in general relativity \cite{lollhttps://doi.org/10.48550/arxiv.2206.06762}. We show how these basic principles combined  lead to the notion of ``relativity'' of quantum superpositions and to invariance of probabilities in quantum mechanics under quantum transformations between classical coordinates, i.e.\ transformations between sets of coordinates associated with different amplitudes of a system in a superposition. This is related to but is conceptually simpler than and distinct from the approach of quantum reference frames \cite{giacomini2019,giacomini2021einsteins,giacomini2021quantum,Giacomini2021spacetimequantum,kabelhttps://doi.org/10.48550/arxiv.2207.00021,delahamettehttps://doi.org/10.48550/arxiv.2112.11473,Castro-Ruiz2020,delaHamette2020quantumreference}. We also note that while studies using EFT approaches to quantum gravity at low energies have successfully been carried out in a variety of contexts \cite{donoghuePhysRevD.50.3874,blencowePhysRevLett.111.021302,huPhysRevD.58.125021,goldbergerPhysRevD.73.104029,donoghue10.1063/1.4756964,arteagaPhysRevD.70.044019}, the focus of these studies (typically the dynamics of spin-2 gravitons propagating on a pre-existing background spacetime) is distinct from the issue we address in this paper, i.e.\ superpositions of semiclassical states of gravitational fields. 

Although many claims have been made that such scenarios represent genuine examples of a quantum-gravitational metric \cite{christodoulou2019possibility,christodolouhttps://doi.org/10.48550/arxiv.2202.03368,fragkoshttps://doi.org/10.48550/arxiv.2206.00558,ligezPhysRevLett.130.101502,Chen2023quantumstatesof,belenchiadoi:10.1142/S0218271819430016} (quantum-gravitational insofar as such solutions cannot be described using the frameworks provided by quantum theory and general relativity), or in the nonrelativistic limit, superpositions of Newtonian gravitational fields \cite{Carlesso_2019,danielsonPhysRevD.105.086001,marlettoPhysRevD.98.046001,overstreetPhysRevD.108.084038,Anastopoulos_2015,Anastopoulos_2020,CARNEYPhysRevD.105.024029,mikiPhysRevD.103.026017,matsamuraPhysRevA.106.012214,kakuPhysRevD.106.126005}, our result shows that this is not unambiguously true. Just as for classical configurations, where only relative distances between objects are of physical significance, the same is also true for gravitational sources in a superposition of different configurations, where these configurations are related by a coordinate transformation \cite{zych2018relativity}. 

To understand the impact of the above result, we apply it in four distinct contexts of recent interest: gravitationally-induced entanglement~\cite{bose2017spin,marletto2017gravitationally,christodoulou2019possibility}, decoherence of spatial superpositions of black holes \cite{arrasmith2019,Demers_1996}, and in the Appendices, decoherence of dark matter~\cite{allaliPhysRevLett.127.031301}, and the gedanken experiment in Ref.\ \cite{belenchiaPhysRevD.98.126009}.\ We discuss the implications of our findings for the conclusions that can be drawn from such scenarios and possibilities for lifting the inherent ambiguities by considering superpositions of non-diffeomorphic metrics.

\section{Preliminaries}

\subsection{Relativity of Quantum Superpositions}\label{sec:IIArelativity}
Before discussing  superpositions of gravitating quantum systems, it is important to clarify the key concepts relevant to that discussion in the general context of superpositions of quantum states, without including any interactions.

General covariance demands that absolute position has no physical meaning \cite{Weinberg1972-WEIGAC}. For example, we may consider a scenario with two particles, one localized at some position $x$ and another at some translated position labeled as $x + X$. In any translationally invariant theory, this scenario is equivalent to the scenario in which the first particle is assigned a position $x - X$ and the second one is assigned position  $x$. Any difference between the two scenarios is merely apparent, attributed to a choice of a coordinate system, which emphasises that only relative distances are physically meaningful, with coordinate systems playing a subordinate, rather than fundamental role.   

We now outline why this idea, despite intuition, does extend to quantum theory, which we then present rigorously in the next section in the context of superpositions of ``semiclassical'' (i.e.\ coherent) states of spacetime.

Consider a scenario with two particles, where the first is assigned a state $|x \rangle$ corresponding to a position $x$, while the second is in a superposition of positions, 
\begin{align}
    \frac{1}{\sqrt{2}} ( | x + X_1 \rangle + | x + X_2 \rangle ), 
    \nonumber 
\end{align}
where $\langle x + X_1 | x + X_2 \rangle = 0$. If indeed only relative distances are physically meaningful in describing the configuration of the particles, then the joint state
\begin{align}
    \frac{1}{\sqrt{2}} | x_1 \rangle \otimes ( | x_2 + X_1 \rangle + | x_2 + X_2 \rangle ) 
    \label{eq1state}
\end{align}
would be equivalent to a seemingly different state 
\begin{align}
    \frac{1}{\sqrt{2}} ( | x_1 - X_1 \rangle + | x_1 - X_2 \rangle ) \otimes | x_2 \rangle  
    \label{eq2state}
\end{align}
in which the second particle is localized at a position $x$, while the first is in a superposition of translations. 
For each individual amplitude, the two states can be related by a passive translation, represented as a unitary operator 
\begin{align}
    \hat{T}(X_i) \equiv \hat T_1( X_i ) \otimes \hat T_2 (X_i)
    \nonumber \vt 
\end{align}
where $i = 1,2,$ that enacts a change of classical coordinates labelling the positions, and thus relabelling the basis states of all the systems (here, the two particles): 
\begin{align}
    | x_1 \rangle \otimes | x_2 + X_i \rangle &= \hat{T} (X_i) | x_1 - X_i \rangle \otimes | x_2 \rangle  . 
    \vt 
\end{align}
The existence of such a transformation is a consequence of the linearity of quantum mechanics i.e.\ Wigner's theorem \cite{wigner1931gruppentheorie}. Importantly, a mapping between the superposition states, denoted generically by the operator $\hat{\mathcal{T}}$, also exists:
\begin{align}\label{eq1}
    & \frac{1}{\sqrt{2}}| x_1 \rangle ( | x_2 + X_1 \rangle + |x_2 + X_2 \rangle ) \nonumber 
    \\
    & \qquad = \frac{\hat{\mathcal{T}}}{\sqrt{2}} ( |x_1 - X_1 \rangle + |x_1 - X_2 \rangle) | x_2 \rangle,
\end{align}
where $\hat{\mathcal{T}}$ in the present case takes the particularly simple form, 
\begin{align}
    \hat{\mathcal{T}} &= \sum_{i=1,2} \hat{T}(X_i) (  | x_1 - X_i \rangle\langle x_1 - X_i | ),
    \nonumber 
\end{align}
which is essentially an application of the translation operator $\hat{T}(X_i)$, quantum-controlled on the basis states associated with particle $1$. While it is uncontroversial that such an operator $\hat{\mathcal{T}}$ exists and can be seen as change of basis in the two-particle Hilbert space, we argue in the next section that it can also be seen as a quantum change of coordinates -- to coordinates that are ``in a superposition'' of different translations relative to the original coordinate system.

\begin{figure}[h]
    \centering
    \includegraphics[trim={25cm 0 25cm 0}, width=\linewidth]{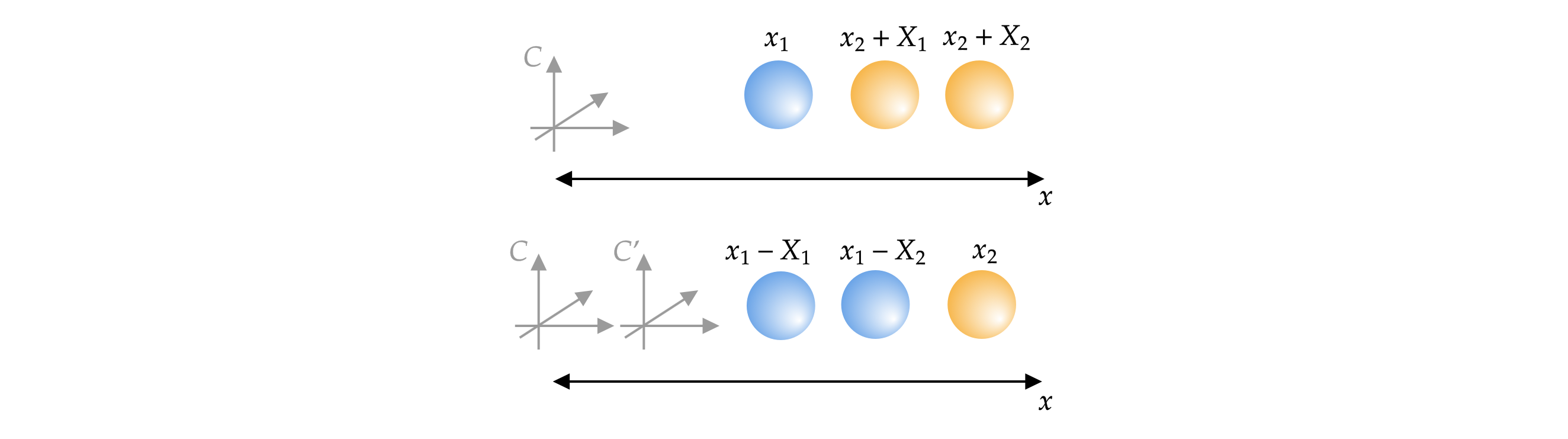}
    \caption{Schematic diagram illustrating the states Eq.\ (\ref{eq1state}) and (\ref{eq2state}) using two sets of quantum coordinates. In the top panel, particle 1 is a fixed distance $x_1$ from the origin of coordinates (denoted $C$), while particle 2 is in a superposition of two positions $x_2 + X_1, x_2 + X_2$. In the bottom panel, the coordinate system is ``in superposition'' relative to $C$ displayed in the top panel, but relative distances between the two particles are preserved in each branch.}
    \label{fig:qrf}
\end{figure}

We stress that what Eq.\ (\ref{eq1}) represents is that the choice and labelling of basis states for a quantum system is merely conventional, and  we  recognise that this freedom to relabel the basis is represented by a unitary mapping between different bases, even if no classical coordinate transformation exists that such a unitary represents. We sketch an example of the above discussed interpretation of the unitary implied by Eq.~\eqref{eq1} as a transformation to coordinates in a ``superposition'' of translations in Fig.\ \ref{fig:qrf}, for the case of two particles in an entangled state and separated by the same distance $X$ in each amplitude.

\vspace{10pt}

\subsection{Relativity of Spacetime Superpositions}\label{sec2b}

\subsubsection{Hilbert Space and Semiclassical Spacetime States}\label{sec:hilbertspaceassumptions}
In dealing with quantum superpositions of spacetime and dynamics of additional  DoFs residing within such spacetimes, we adopt standard assumptions used in quantum gravity phenomenology. In particular:
\begin{enumerate}
    \item The gravitational field can be placed in a superposition of ``semiclassical'' states, $| g(x) \rangle, | g(y) \rangle$, each corresponding to coherent states of the metric parametrized by coordinates $x,y$ and associated with a classical solution to Einstein's field equations for appropriate matter distributions.
    \item The dynamics of other DoFs in the presence of the superposition of spacetimes 
    can be described by quantum-controlled dynamics: in each of the amplitudes of the superposition the dynamics are described by quantum theory in the spacetime of that amplitude.
    \item The Hilbert space of the combined spacetime and system DoFs can be described as the tensor product $\mathcal{H} = \mathcal{H}_\mathrm{S} \otimes \mathcal{H}_\mathrm{DoF}$ where, depending on the context, $\mathcal{H}_\mathrm{DoF}$ can be further decomposed as a product of different matter and/or field DoFs (for example when considering the light-matter interaction, see Sec.\ \ref{secIIe}). 
\end{enumerate}

\subsubsection{Unitary Representation of Diffeomorphisms}\label{secIIB2}

\noindent A (global) diffeomorphism between the manifolds $M, M'$ is a smooth map $\varphi: M \mapsto M'$ with smooth inverse map $\varphi^{-1}$. The manifolds $M,M'$ represent the same geometry and are therefore physically indistinguishable. To emphasise this, we denote two diffeomorphically related manifolds each endowed with a metric $g(x), g(y)$ with the same ``$g$'' but parametrised by different coordinate systems $x,y$. The coordinates $x,y$ are associated with the respective configurations of the source mass, and could describe for example coordinates centered at the position of the mass. Since diffeomorphisms in general relativity represent a symmetry transformation, they can be represented by unitary operators acting on the coordinates \cite{Loveridge2018}. Therefore, the semiclassical configurations associated with the states $| g(x) \rangle, | g(y) \rangle$ are related via the action of the unitary operator $\hat{T}(x,y)$,
\begin{align}
    | g(x) \rangle &= \hat{T}(x,y) | g(y) \rangle  
    \label{eq7}
\end{align}
and its inverse, 
\begin{align}
    | g(y) \rangle &= \hat{T}^{\dd}(x,y) | g(x) \rangle 
    \label{eq8transformation}
\end{align}
where $\hat{T}(x,y)$ could depend on some potentially complicated function of the spacetime coordinates that characterises the distance between the points in $x,y$ respectively. As an illustration, consider the case where $| g(x) \rangle, | g(y) \rangle$ are related by a translation of the coordinates $X$. Then, we can write
\begin{align}
    | g(x) \rangle &= \hat{T}(X) | g(y) \rangle .
    \vphantom{\frac{1}{\sqrt{2}}} 
    \nonumber 
\end{align}

\subsubsection{Initial and Final States}\label{secIIB3}

Throughout this paper we will be interested in the dynamics of quantum systems residing within the diffeomorphically related spacetime superpositions discussed so far. For example, we will analyze initial states of the form, 
\begin{align}
    \frac{1}{\sqrt{2}} ( | g(x) \rangle + | g(y) \rangle ) \otimes | \phi_\mathrm{DoF} \rangle 
    \label{eq10initialstate}
\end{align}
describing the case where the state of the additional DoFs $| \phi_\mathrm{DoF}\rangle$ factors from the state of the metric, as well as entangled states
\begin{align}
    \frac{1}{\sqrt{2}} ( | g(x) , \phi(x) \rangle + | g(y) , \phi(y) \rangle ) 
    \nonumber 
\end{align}
where the shorthand notation $| g(\cdot) , \phi(\cdot) \rangle \equiv | g( \cdot) \rangle | \phi_\mathrm{DoF} (\cdot) \rangle\equiv  | g( \cdot) \rangle\otimes | \phi_\mathrm{DoF} (\cdot) \rangle$ is implied. In certain cases, we will examine a specific interaction between the additional DoFs, such as that of a two-level system (modelling some matter DoFs) with a quantum field, for which  we employ the explicit notation, 
\begin{align}
    | \phi_\mrm{DoF} \rangle &= | \phi_\mathrm{M} \rangle 
    | \phi_\mathrm{F}  \rangle .
    \nonumber 
    \vt 
\end{align}
To denote the final state of all involved systems, we use the state vector $| \Omega \rangle$. In some contexts we also look at conditional states given a measurement on only some subsystems. In this case we use the same subscripts as for initial states to label the different DoFs.

\subsubsection{Unitary Transformations of Quantum Coordinates Acting on Multiple DoFs}  

We are interested in scenarios combining the concepts of Sec.\ \ref{secIIB2} and \ref{secIIB3}. For example, consider the initial state Eq.\ (\ref{eq10initialstate}). As stated in Eq.\ (\ref{eq7}), the operator $\hat{T}(x,y)$ will act on the coordinates of all the systems, which here means both the metric \textit{and} any additional DoFs residing within. Implicitly then, $\hat{T}(x,y) \equiv \hat{T}_\mathrm{G}(x,y) \otimes \hat{T}_\mathrm{DoF}(x,y)$ where $\hat T_\mathrm{G}(x,y)$ acts on the metric Hilbert space. As such, Eq.\ (\ref{eq10initialstate}) may be equivalently represented as (dropping the normalisation factor for simplicity), 
\begin{align}
    & | g(x) \rangle | \phi_\mathrm{DoF} \rangle + \hat{T}^\dd(x,y) | g(x) \rangle |{\hat{T}_\mathrm{DoF}(x,y)}  \phi_\mathrm{DoF} \rangle 
    \nonumber 
    \VT 
    \\
    & \qquad = | g(x) \rangle | \phi_\mathrm{DoF} \rangle + \hat{T}^\dd_\mathrm{G} | g(x) \rangle \otimes \hat{T}_\mathrm{DoF}^\dd | \hat{T}_\mathrm{DoF} \phi_\mathrm{DoF} \rangle 
    \VT \label{eq8nomenclature}
\end{align}
having dropped the coordinates $(x,y)$ in the second line for simplicity. We have used the notation $| \hat{T}_\mathrm{DoF} \phi_\mathrm{DoF} \rangle \equiv \hat T_\mrm{DoF} | \phi_\mrm{DoF} \rangle$ to denote the inverse of the transformation $\hat{T}^\dd_\mathrm{DoF}$, such that 
\begin{align} 
    \hat{T}^\dd_\mathrm{DoF} | \hat{T}_\mathrm{DoF} \phi_\mathrm{DoF} \rangle = | \phi_\mathrm{DoF} \rangle . 
\vphantom{\frac{1}{\sqrt{2}}} 
\nonumber 
\end{align}

\subsection{Examples}

\subsubsection{Post-Newtonian Metric in Isotropic Coordinates}

As a concrete application of the ideas presented above, consider the post-Newtonian expansion of the Schwarzschild metric in isotropic coordinates $(t,\rho, \theta, \phi)$, 
\begin{align}
    \D s^2 &= - \bigg( 1 + \frac{2 \Phi( \rho - \rho_0 ) }{c^2} \bigg)\D t^2 
    \nonvt \\
    & + \bigg( 1 - \frac{2 \Phi ( \rho - \rho_0 ) }{c^2} \bigg) \big( \D \rho^2 + ( \rho- \rho_0)^2 \D \Omega^2 \big) 
    \vt 
    \label{eq9}
\end{align}
where $\D \Omega^2 = \D \theta^2 + \sin^2(\theta) \D \phi^2$, $\Phi \equiv \Phi ( \rho - \rho_0)$ is the gravitational potential, and $\rho_0$ is a displacement of the source mass from the origin. One might suppose that the source mass is located in a superposition of the following coordinates relative to a test particle,
\begin{align}
    (t, \rho_0, \theta_0, \phi_0 )  ,  ( t , \rho_0', \theta_0, \phi_0 ) 
    \nonumber
    \vt 
\end{align}  
and so one can write the joint state as (up to normalisation), 
\begin{align}
    ( | g ( \rho_0) \rangle + | g ( \rho_0' ) \rangle ) \otimes | \phi_\mrm{test} \rangle 
    \vt 
    \label{eq10}
\end{align}
Since $\rho_0, \rho_0'$ are related by a coordinate transformation generated by the unitary $\hat T^\dd$ i.e.\ $| g ( \rho_0') \rangle = \hat T_\mrm{G}^\dd( \rho_0 , \rho_0') | g( \rho_0) \rangle$, one can write this state in the form, 
\begin{align}
     | g ( \rho_0) \rangle | \phi_\mrm{test} \rangle + \hat T_\mrm{G}^\dd | g( \rho_0) \rangle \otimes \hat T_\mrm{test}^\dd | \hat T_\mrm{test} \phi_\mrm{test} \rangle 
     \vt 
     \label{eq11}
\end{align}
in a similar manner to Eq.\ (\ref{eq8nomenclature}).

\subsubsection{De Sitter Spacetime in Static Coordinates}

As another example, let us consider de Sitter spacetime, described by the hyperboloid, 
\begin{align}
    - Z_0^2 + Z_1^2 + Z_2^2 + Z_3^2 + Z_4^2 &= a^2
    \vt 
\end{align}
where $a = \sqrt{\Lambda/3}$ with $\Lambda$ the cosmological constant, embedded in a flat five-dimensional Minkowski spacetime, 
\begin{align}
    \D s^2 &= - \D Z_0^2 + \D Z_1^2 + \D Z_2^2 + \D Z_3^3 + \D Z_4^2 
    \vt 
\end{align}
and where $\Lambda$ is the cosmological constant. The spherically symmetric coordinates $(T,R, \theta, \phi)$ can be introduced by taking,
\begin{align}
    Z_0 = Z_0' &= \sqrt{a^2 - R_0^2} \sinh( T/a ) 
    \nonvt \\
    Z_1 = Z_1' &= \sqrt{a^2 - R_0^2} \cosh (T/a) 
    \nonvt \\
    Z_2 = R_0 \cos(\theta_0) , Z_2' &= R \cos(\theta_0') 
    \nonvt \\
    Z_3 = R_0 \sin(\theta) \cos(\phi) , Z_3' &= R_0 \sin(\theta_0') \cos(\phi_0') 
    \nonvt \\
    Z_4 = R_0 \sin(\theta_0) \sin(\phi_0) , Z_4' &= R \sin(\theta_0') \sin(\phi_0') 
    \vt 
\end{align}
where the unprimed embedding coordinates $(Z_i)$ represent one branch of the superposition, while the primed coordinates $(Z_i')$ represent the embedding coordinates of the other branch of the superposition. The two coordinate systems thus differ by rotations in the angular $(\theta)$ and azimuthal $(\phi)$ planes. In direct analogy to the prior example, one can write down similar dual representations of the metric and a test particle state as in Eqs.\ (\ref{eq10}) and (\ref{eq11}).

\section{Relativity of Spacetime Superpositions}
\label{relspacesup}

\subsection{Transition Amplitudes 
}\label{sec:IIIA}

Let us now apply the introduced assumptions and formalism to the problem of matter sourcing the metric that is in a superposition of diffeomorphically related states. We emphasise that the results presented here depend only on the linearity of quantum theory (insofar as symmetry transformations such as translations are described by unitary operators) and the invariance of physical laws under ordinary coordinate transformations. First consider a source mass and the corresponding metric in a superposition of classically distinct configurations  $| g(x) \rangle, |g (y) \rangle$, described by the state
\begin{align}\label{eq2}
    | \psi \rangle &= \frac{1}{\sqrt{2}} \left( | g(x) \rangle + | g(y) \right)
\end{align}
(we generalize this to arbitrary superposition states in the Appendix). It is helpful here to conceptualise the mass configuration as being quantum-controlled by an ancillary system that can be prepared and measured in appropriate states, in analogy with the standard approach in quantum measurement theory. That is, for each state of the control there is an associated mass configuration with a classical manifold and gravitational field, relative to the other quantum DoFs. This is a key assumption that underpins numerous recent investigations in the area of spacetime superpositions, including analyzes of gravitationally-induced entanglement proposals \cite{bose2017spin,marletto2017gravitationally}, spacetime quantum reference frames \cite{giacomini2021einsteins,giacomini2021quantum,Giacomini2021spacetimequantum}, and decoherence \cite{arrasmith2019,Demers_1996,allaliPhysRevLett.127.031301}.

According to Eq.\ (\ref{eq8transformation}), we can express the spacetime superposition state above as
\begin{align}
   | \bar{\psi} \rangle &= \frac{1}{\sqrt{2}} ( | g(x) \rangle + \hat{T}^\dd(x,y) | g(x) \rangle ) 
    \nonumber 
    \\
    &= \frac{1}{\sqrt{2}} ( \hat I + \hat{T}^\dd(x,y) ) | g(x) \rangle . 
    \label{eq16transformationequivalence}
\end{align}
where by construction $| \bar{\psi} \rangle \equiv | \psi \rangle$ but we use this  notation to distinguish the \textit{representation} of the state from that in Eq.~(\ref{eq2}), though as emphasised, the two states are equivalent. We reiterate that this relationship follows from general relativity and the basic tenets of quantum theory, namely that symmetries can be represented with unitary operators. We also stress that in the case of translations, the spacetimes are related by a simple relabelling of, for example, the origin of coordinates, and not an active translation of the source mass through some pre-existing curved spacetime. 

Let us now consider some additional DoF in the state $| \phi_\mathrm{DoF} \rangle$, and a joint initial state of the form,
\begin{align}\label{eq4}
    | \psi \rangle &= \frac{1}{\sqrt{2}} \left( | g(x) \rangle + | g(y) \rangle \right) | \phi_\mathrm{DoF} \rangle .
\end{align}
where as discussed, the physical system represented by the state $| \phi_\mathrm{DoF} \rangle$ could be, for example, a quantum field or a particle in first-quantization, or both of these interacting with each other--we will consider all of these in the applications of our approach in the next sections and the Appendix. The only assumption behind Eq.~\eqref{eq4} is that the state of the matter DoFs is uncorrelated with that of the source mass (and thus of the spacetime), but apart from this, is arbitrary. Again, the assumption of an uncorrelated initial state is not necessary, and indeed we will also consider correlated initial states in Sec.\ \ref{sec:apparent}.

Let us 
consider time evolution of the systems--including free dynamics as well as possible interactions between them--denoted by $\hat U$. In general, the time evolution of the matter DoFs can depend on the state of the source mass, and so the operator $\hat U$ can be represented as
\begin{align}
    \label{eq12evolutionoperator}
    \hat{U} = \hat{U}(x) \otimes | g(x) \rangle\langle g(x) | + \hat{U} (y) \otimes | g(y) \rangle\langle g(y) | ,  
\end{align}
where $\hat{U}(x) $ and $\hat{U}(y)$ individually govern the time-evolution of all the DoFs on the respective spacetime. The quantum-controlled evolution operator in Eq.\ (\ref{eq12evolutionoperator}) is consistent with the assumptions introduced in Sec.\ \ref{sec:hilbertspaceassumptions}, in the sense of a low-energy limit. We note that operators of this form are ubiquitous in, for example, quantum information theory--see for e.g.\ Refs.\ \cite{nielsen2010quantum,eblerPhysRevLett.120.120502,Chiribella_2021njp,rubinoPhysRevResearch.3.013093,blondeauPhysRevA.104.032214}. Moreover, we note that Eq.\ \eqref{eq12evolutionoperator} assumes that the free evolution of the individual metric states $| g(x) \rangle, |g(y) \rangle$ is trivial. While this assumption holds in the examples considered, our formalism directly applies also to cases where the dynamics of the metric directly follows that of the source matter (as in the argument of Ref.\ \cite{christodoulou2019possibility} in Sec.\ \ref{sec:GIEsec2}).

Denoting $| \Omega_\mrm{G} \rangle$ a final state of the gravitational DoFs, and by $| \Omega_\mathrm{DoF} \rangle$ the state of the matter DoFs, the general form of the probability amplitude reads
\begin{align}
    &\langle \Omega_\mrm{DoF} |  \langle \Omega_\mrm{G} | \hat{U} | \psi \rangle = \frac{1}{\sqrt{2}}  ( \langle \Omega_\mrm{DoF} | \langle \Omega_\mrm{G} | \hat{U}(x)  | g(x) \rangle 
    \nonumber \\
    & \:\:\: + \langle \Omega_\mrm{DoF} | \langle \Omega_\mrm{G} | \hat{U}(y) | g(y) \rangle  ) | \phi_\mathrm{DoF} \rangle 
    \vt 
    \label{eq19}
\end{align}
We wish to enact a transformation similar  to (\ref{eq16transformationequivalence}) on the joint state of the metric and additional DoFs.  Since changing the coordinates transforms the states of all involved systems, it acts both on the metric $g(y)$ as well as on the additional DoFs $\phi_\mrm{DoF}$ in that branch of the superposition. That is, $\hat{T}(x,y) = \hat{T}_\mrm{G}(x,y) \otimes \hat{T}_\mrm{DoF}(x,y)\equiv \hat{T}_\mrm{G} \otimes \hat{T}_\mrm{DoF}$. Equation (\ref{eq19}) thus becomes, 
\begin{align}
    & \langle \Omega_\mrm{DoF} | \langle \Omega_\mrm{G} | \hat{U} | \psi \rangle 
    = \frac{1}{\sqrt{2}} \langle \Omega_\mrm{DoF} | \langle \Omega_\mrm{G} | \hat{U}(x)| g(x) \rangle | \phi_\mrm{DoF} \rangle 
    \nonumber 
    \\
    & \:\:\: + \frac{1}{\sqrt{2}} \langle \Omega_\mrm{DoF} | \langle \Omega_\mrm{G} | \hat{U}(y) \hat{T}^\dd | g(x) \rangle | \hat{T}_\mrm{DoF} \phi_\mrm{DoF} \rangle 
    \nonumber 
\end{align}
having dropped the explicit dependence on the coordinates $x,y$ for brevity. Inserting the identity operator $\hat I = \hat{T}^\dd \hat{T}$ in the second term gives, 
\begin{align}\label{eq8diffeo}
    &\langle \Omega_\mrm{DoF} | \langle \Omega_\mrm{G} | \hat{U} | \psi \rangle = \frac{1}{\sqrt{2}} \langle \Omega_\mrm{DoF} |\langle \Omega_\mrm{G} | \hat{U}(x) |g(x) \rangle| \phi_\mrm{DoF} \rangle 
    \non \\
    & \:\:\: + \frac{1}{\sqrt{2}} \langle \hat{T}_{\mrm{DoF}} \Omega_\mrm{DoF} | \langle \hat{T}_{\mrm{G}}\Omega_\mrm{G} | \hat{T} \hat{U}(y) \hat{T}^\dd | g(x) \rangle | \hat{T}_\mrm{DoF} \phi_\mrm{DoF} \rangle , 
\end{align}
where the last line follows from $\langle \hat{T}_\mrm{(\cdot)} \Omega_\mrm{(\cdot)} | = \langle \Omega_\mrm{(\cdot)} |\hat{T}_\mrm{(\cdot)}^\dd$. 
Importantly, as long as $\hat{T} \equiv \hat{T}(x,y)$ enacts a diffeomorphism, then
\begin{align} 
    \hat{T} \hat{U}(y) \hat{T}^\dd \equiv \hat{U}(x)
    \vphantom{\frac{1}{\sqrt{2}}} 
 \label{TUT}
\end{align}
which amounts to enacting a transformation of coordinates on the evolution operator on the second branch of the superposition, and thus we are left with the amplitude
\begin{align}\label{eq8diffeo_2}
    & \langle \Omega_\mrm{DoF} | \langle \Omega_\mrm{G} | \hat{U} | \psi \rangle =\frac{1}{\sqrt{2}} \langle \Omega_\mrm{DoF} | \langle \Omega_\mrm{G} | \hat{U}(x) | g(x) \rangle  | \phi_\mrm{DoF} \rangle 
    \non 
    \\
    & \:\:\: + \frac{1}{\sqrt{2}} \langle \hat{T}_\mrm{DoF} \Omega_\mrm{DoF} | \langle \hat{T}_{\mrm{G}}\Omega_\mrm{G} | \hat{U}(x) | g(x) \rangle | \hat{T}_\mrm{DoF} \phi_\mrm{DoF} \rangle .
\end{align}
Equations~\eqref{eq19} and \eqref{eq8diffeo_2} describe the same scenario--they demonstrate that the same dynamics can be interpreted as taking place in spacetime that is in superposition of states $|g(x)\rangle, |g(y)\rangle $ in Eq.~\eqref{eq19}, or in a single spacetime described by the state $|g(x)\rangle$, where now all the preparations and measurements are in superposition, as expressed in Eq.~\eqref{eq8diffeo_2}. We remark that this analysis follows analogously even if the gravitating source mass is in a spatial superposition of extended wavepackets. So long as one can relate the amplitudes of the superposition via an appropriate unitary transformation, the conclusions found here stand.

\subsection{Full Interferometric Scenario}\label{sec:IIIB} 


We now apply the above to a generic interferometric scenario as this is most relevant to recent literature.
In particular, consider the representation Eq.\ (\ref{eq19}), 
where 
the source mass/spacetime is measured in the superposition basis, including the state $| \Omega_\mrm{G} \rangle = ( | g(x) \rangle + | g(y) \rangle )/\sqrt{2}$. For mutually orthogonal source mass/metric states, the final amplitude reads
\begin{align}\label{eq:amplitude1}
    \frac{1}{2}  \langle \Omega_\mrm{DoF} | ( \hat{U}(x) + \hat{U}(y) ) | \phi_\mrm{DoF} \rangle .
\end{align}
The interpretation of Eq.~\eqref{eq:amplitude1} is that matter DoFs prepared in the state $|  \phi_\mrm{DoF}\rangle$  evolve in a coherent superposition of ``paths'' associated with the different amplitudes of the spacetime superposition. 

On the other hand, if we adopt the representation given by Eq.~\eqref{eq8diffeo_2} the same amplitude equivalently reads 
\begin{align}\label{eq12}
    & \frac{1}{2} \langle \Omega_\mrm{DoF} | \hat{U}(x) | \phi_\mrm{DoF}\rangle 
    \nonumber 
    \VT \\
    & + \frac{1}{2} \langle \hat{T}_\mrm{DoF} \Omega_\mrm{DoF} | \hat{U}(x) | \hat{T}_\mrm{DoF} \Omega_\mrm{DoF} \rangle ,
\end{align}
where we used the fact that the two mass configurations and metrics are macroscopically distinct, meaning 
\begin{align}
    \langle g(x) | \hat{T}_\mrm{G}^\dd | g(x) \rangle = \langle g(x) | \hat{T}_\mrm{G} | g(x) \rangle = 0 
    \VT 
    \nonumber 
\end{align}
and again that $\hat{T} \hat{U}(y) \hat{T}^\dd \equiv \hat{U}(x)$.

The fact that Eqs~\eqref{eq:amplitude1} and \eqref{eq12} are the same probability amplitude is a formal expression of the fact that a scenario involving quantum systems in a spacetime sourced by a spatial superposition of mass configurations is  equivalent to a scenario where the particle follows different superposed trajectories in one spacetime and is measured as such. The physical interpretation of the probability amplitude is thus inherently ambiguous. The equivalence between the two scenarios can be interpreted as describing the same physical situation  using two sets of coordinates that are related  by a ``superposition'' of classical transformations \cite{zych2018relativity}.
The equivalence of amplitudes further implies that probabilities are identical in both scenarios. 

This is a key point of this article and one that we believe is currently overlooked. ``Quantum superpositions of spacetimes'' can therefore be ambiguous, as in the example given here, as they can be re-expressed in terms of modified initial states and measurements of the remaining DoFs, but where the the source mass, and thus the spacetime, can be treated as classical and remains fixed. This equivalence between representations is especially important given the recent interest in identifying quantum-gravitational effects that may arise from spatial superpositions of a source mass \cite{dewitt2011role}. 

\begin{figure}[h]
    \centering
    \includegraphics[trim={20cm 0 20cm 0}, width=\linewidth]{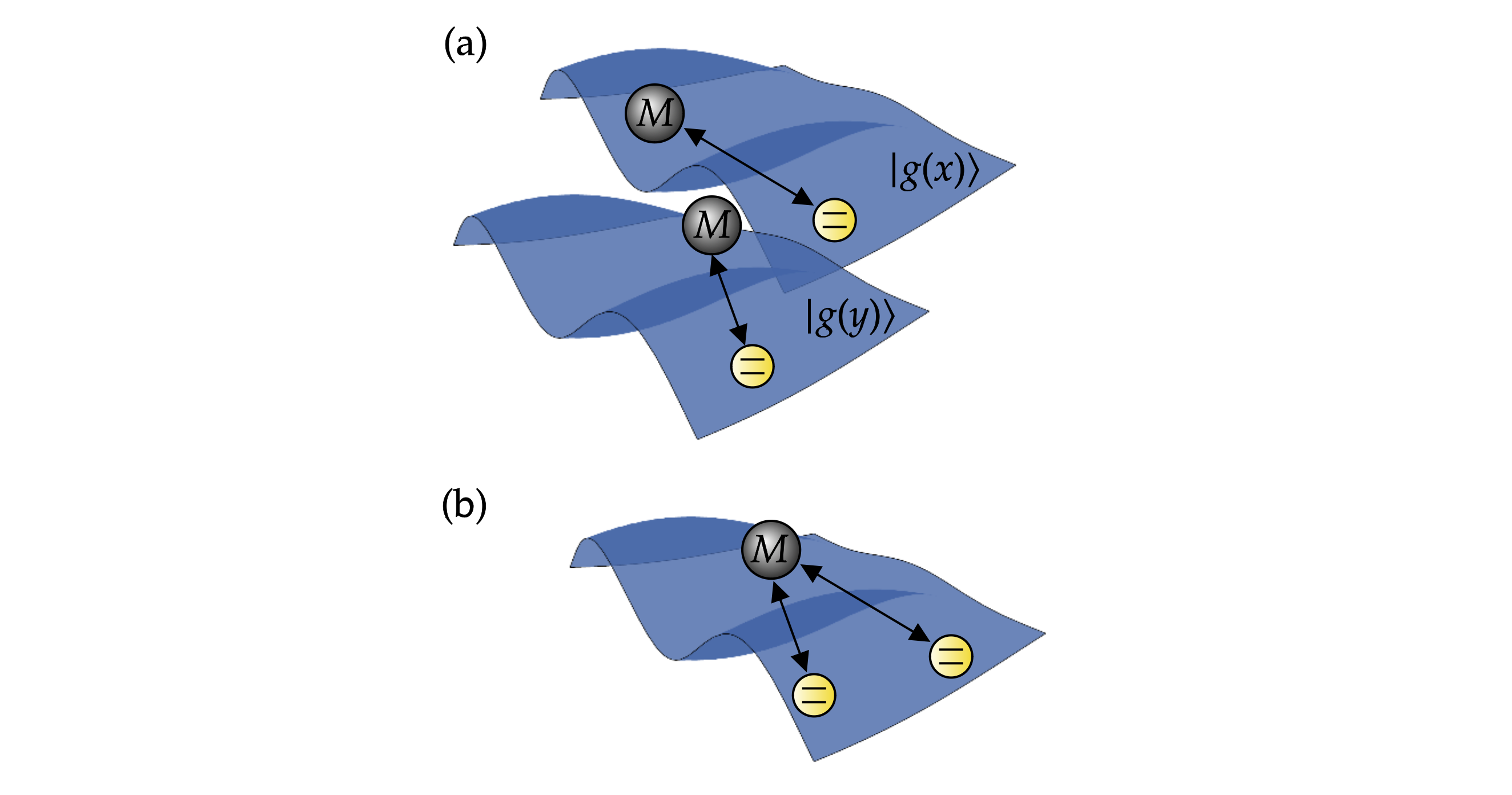}
    \caption{Illustration of how the same scenario viewed from different coordinates acquires the interpretation of (a) superposition of spacetimes,  (b) superposition of locations of matter DoFs (e.g. a test particle) with respect to one mass configuration. The curved ``background'' is illustrative only.}
    \label{fig:diffeomorphism}
\end{figure}

The crux of the argument is that only \textit{relative configurations} between the interacting systems are physically relevant, such as a superposition of two distances between a pair of particles; a global (joint) coordinate transformation enacted on all degrees of freedom is irrelevant. Scenarios where quantum probes are situated on a classical background while the time-evolution occurs in a superposition of trajectories were recently considered in Refs.\ \cite{foo2020unruhdewitt,fooPhysRevResearch.3.043056,fooPhysRevD.103.065013,Foo_2021,barbadoPhysRevD.102.045002}. For illustration,  we have sketched schematically in Fig.\ \ref{fig:diffeomorphism} such scenario where the superposed amplitudes differ by a translation.

Let us further take an approximation that the two amplitudes of the matter
DoFs in Eq.~\eqref{eq12} are orthogonal throughout the evolution, i.e.~$\langle \phi_\mrm{DoF} |\hat{T}_\mrm{DoF}\phi_\mrm{DoF}\rangle = 0$ and evolve only by a phase -- which is also made in the gravitationally-induced entanglement proposals of Refs.\ \cite{bose2017spin, marletto2017gravitationally}. Here this means that $|\Omega_\mrm{DoF}^+\rangle:=(| \phi_\mrm{DoF} \rangle  + e^{i\varphi}| \hat{T}_\mrm{DoF}\phi_\mrm{DoF}\rangle) /\sqrt{2} $ and defining analogously $|\phi_\mrm{DoF}^+\rangle:=(| \phi_\mrm{DoF} \rangle  + |\hat{T}_\mrm{DoF}\phi_\mrm{DoF} \rangle) /\sqrt{2} $ we obtain
that Eq.~\eqref{eq12} can be written as
\begin{equation}    \langle\Omega_\mrm{DoF}^+|\hat{U}(x)|\phi_\mrm{DoF}^+\rangle.
\end{equation}
This means that the scenario (a) in which the source mass is prepared and measured in superposition while the other DoFs evolve in the resulting superposition of metrics is physically equivalent and thus indistinguishable to the scenario (b) in which the other DoFs are prepared and measured in superposition while the source mass remains in some fixed  state sourcing a classical metric. 

We conclude this section by mentioning that the formalism presented here has a well-defined nonrelativistic limit both in terms of DoFs on the metric as well as source matter and metric. Indeed, our approach inherits the standard understanding of measurements in quantum field theory (e.g.\ spacetime smearing of position operators), which in the nonrelativistic limit is well-described by the position eigenstates utilized in Sec.\ \ref{sec:IIArelativity}.
In terms of source matter and metric,  it can describe matter  that sources a Newtonian potential and through this potential interacts with other massive particles. This is inherited in turn from the fact that the classical metric reduces to the Newtonian description in a low-energy, nonrelativistic limit. For source masses in quantum states such a limit results in ``superpositions of Newtonian potentials,'' which have garnered just as much interest as their general relativistic counterparts (see for example Refs.\ \cite{Carlesso_2019,danielsonPhysRevD.105.086001,overstreetPhysRevD.108.084038,belenchiadoi:10.1142/S0218271819430016,Anastopoulos_2015,Anastopoulos_2020,CARNEYPhysRevD.105.024029,mikiPhysRevD.103.026017,matsamuraPhysRevA.106.012214}). We now turn to specific scenarios of interest as applications of our results, starting from the above mentioned superpositions of Newtonian potentials.

\subsection{Superposition of Geometries in the Gravitationally-Induced Entanglement Proposals}\label{sec:GIEsec2}
The previous analysis demonstrated how in general, spacetime superpositions in which the respective amplitudes are related by a diffeomorphism can be re-expressed in terms of a single background spacetime with the states and measurements of the remaining DoFs being suitably adapted. This motivates us to revisit the conclusions one can draw concerning quantum features of gravity from such scenarios.

In this section, we apply our approach and construction to recent proposals by Bose et.\ al.\ \cite{bose2017spin} and Marletto and Vedral \cite{marletto2017gravitationally}, which have attracted significant interest as presenting possibilities for witnessing the quantization of the gravitational field. These gravitationally-induced entanglement (GIE) setups suggest that observing entanglement between two spatially superposed source masses, interacting gravitationally via the Newtonian potential, is not equivalent to matter interferometry on a fixed background and would provide evidence of the quantum nature of gravity (for example, that entanglement is mediated via gravitons \cite{marshmanPhysRevA.101.052110}). This was originally \cite{bose2017spin,marletto2017gravitationally} argued on the basis that local operations and classical communication cannot generate entanglement (LOCC), and thus under the assumption that the interaction is mediated by a local DoF, any observed entanglement must have arisen because gravity acts as a quantum channel. We also draw attention to a recent interpretation of this argument by Christodolou and Rovelli~\cite{christodoulou2019possibility} who argue that such a setup represents a superposition of genuinely distinct spacetime metrics. This interpretation is the general relativistic generalization of related ones that interpret the GIE system as demonstrating a superposition of Newtonian gravitational fields (i.e.\ originally claimed in Ref.\ \cite{bose2017spin} and developed in later works e.g.\ Refs.\ \cite{danielsonPhysRevD.105.086001,Carlesso_2019,marlettoPhysRevD.98.046001}).

\begin{figure}[h]
    \centering
    \includegraphics[trim={20cm 0 20cm 0},width=1\linewidth]{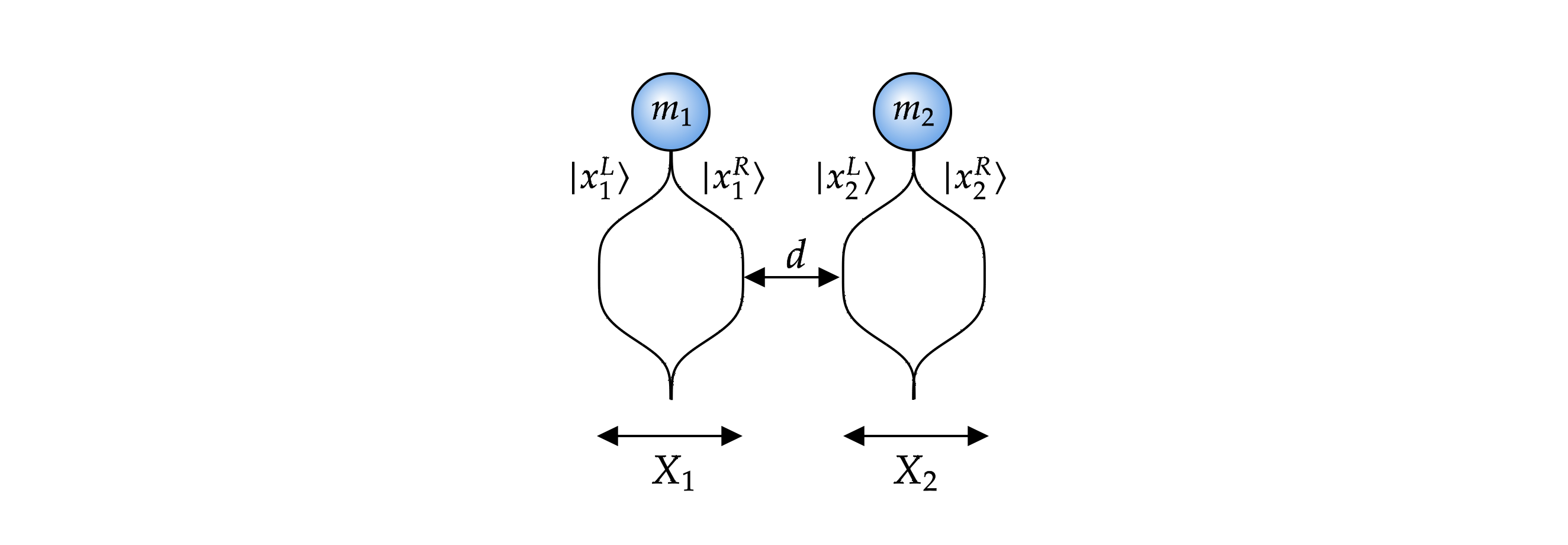}
    \caption{Schematic diagram of the GIE proposal. Each of the states $ | x_i^{L,R} \rangle$ is assumed to be in a highly localised Gaussian wavepacket such that their overlap is zero.}
    \label{fig:gie}
\end{figure}

The GIE proposal is depicted in Fig.\ \ref{fig:gie}. Two masses, $m_1$ and $m_2$, 
are each prepared and measured in spatial superpostion of freely falling trajectories in the uniform gravitational field of Earth (for example via Stern-Gerlach interferometry techniques). Thus initially, each particle is in a state of the form 
\begin{align}
    | \psi_{\mrm{GIE},i} \rangle &= \frac{1}{\sqrt{2}} ( | \psi(x_i^L) \rangle + | \psi(x_i^R) \rangle ) ,  
    \nonumber 
\end{align}
where $ i = 1,2$ labels the particles, while the labels $L$, $R$ denote mutually orthogonal eigenstates of the $x$-component of the spin-1/2 particle, controlling the path that each particle takes. The states $| \psi(x_i^j)\rangle$ are assumed to be highly localised trajectories, and thus we will henceforth denote them as position eigenstates $| x_i^j \rangle$. The initial joint state of the two particles is thus
\begin{align}
    & | \psi_\mathrm{GIE} \rangle = | \psi_{\mathrm{GIE},1} \rangle \otimes | \psi_{\mathrm{GIE,2}} \rangle 
    \nonvt \\
    & \:\:\: = \frac{1}{2} ( | x_1^L , x_2^L \rangle + | x_1^L, x_2^R \rangle 
    + | x_1^R , x_2^L \rangle + | x_1^R , x_2^R \rangle ) 
    \vphantom{\frac{1}{2}} 
    \label{eq22initialstate}
\end{align}
Evolution under mutual gravitational interaction for a time $t$ entangles the test masses by imparting  distance-dependent phases to the components of the superposition. The simplifying assumption that the relative distance in one branch of the superposition state is much smaller than the others (in Fig.\ \ref{fig:gie}, $| x_1^R , x_2^L \rangle$ with relative distance $d$) can be made, yielding the time-evolved state
\begin{align}
    | \psi_\mrm{GIE} \rangle &\to \frac{1}{2} ( | x_1^L , x_2^L \rangle + | x_1^L ,  x_2^R \rangle 
    \nonvt \\
    & \:\:\: + e^{i\phi} | x_1^R , x_2^L \rangle + | x_1^R , x_2^R \rangle ) 
\label{eq22phaseshift}
\end{align}
where $\phi = Gm_1m_2t/\hslash d$ is the phase shift induced by interaction via the Newtonian potential,
\begin{align}
    \hat U  &= \int\D x_1 \D x_2 \: e^{\frac{i Gm_1m_2t}{\hslash| x_1 - x_2| }} | x_1 \rangle\langle x_1 | \otimes | x_2 \rangle\langle x_2 |
    \label{eq27newtonianinteraction}
\end{align}where the projectors $| x_i \rangle\langle x_i |$ act on the Hilbert space of the $i$th particle.

The argument of Ref.\ \cite{christodoulou2019possibility} is that immediately after the superpositions are created, the external metric $| g \rangle$ produced by both particles has not had time to change appreciably, 
\begin{align}
    & |\psi_{\mrm{GIE}}^{(g)} \rangle = \frac{1}{2} | g \rangle ( | x_1^L \rangle + | x_1^R \rangle  )  ( | x_2^L \rangle + | x_2^R \rangle  )
    \non 
    \\
    \label{eq14}
    &= \frac{1}{2} | g \rangle  ( | x_1^L, x_2^L \rangle + | x_1^L , x_2^R \rangle 
    + | x_1^R , x_2^L \rangle + | x_1^R, x_2^R \rangle  ) 
    \vphantom{\frac{1}{2}}
\end{align}We introduced the notation $| \psi_\mrm{GIE}^{(g)} \rangle$ to highlight that this is the argument made in Ref.\ \cite{christodoulou2019possibility}, distinct from Refs.\ \cite{bose2017spin,marletto2017gravitationally} (for which we use the notation $| \psi_\mrm{GIE} \rangle$). Once the gravitational disturbance has had time to propagate, the authors of Ref.\ \cite{christodoulou2019possibility} argue that each branch of the two-particle state will source, in general, a different metric depending on the relative distance between the particles in that branch:
\begin{align}\label{eq15}
    | \psi_{\mrm{GIE}}^{(g)} \rangle &\to \frac{1}{2} ( | g(x_1^L, x_2^L), x_1^L, x_2^L \rangle + | g(x_1^L, x_2^R), x_1^L, x_2^R  \rangle  \nonumber 
    \vt 
    \\
    & + | g(x_1^R, x_2^L), x_1^R, x_2^L \rangle + | g(x_1^R, x_2^R), x_1^R, x_2^R \rangle ) 
    \vt 
\end{align}
where $g(x_1^A,x_2^B)$, $A,B = R,L$ denote the respective two-body metrics sourced jointly by both particles at positions $x_1^A, x_2^B$. They further remark 
that such a scenario no longer represents a semiclassical spacetime but a genuine superposition of metrics. This is true when referring to the metric generated by both particles.

However, despite the fact that Eq.\ (\ref{eq15}) features four (in-principle) different states of the metric, the GIE proposal (nor its generalisation to a superposition of metrics in Ref.\ \cite{christodoulou2019possibility}) does not depend on a  geometry that is sourced by both particles. The results are derived from gravitational interaction \textit{between} the two particles and thus the metric involved can be interpreted as sourced by, say, particle $i = 1$ (this choice being arbitrary), and 
the setup of Fig.\ \ref{fig:gie} can be reinterpreted as a specific four-path interference experiment but on a fixed classical spacetime.

To see this explicitly (i.e.\ that GIE can be re-interpreted as though one of the particles sources a classical metric for the other), let us first consider the Hamiltonian of particle 2 with mass $m_2$, located at coordinate $x_2^{D'}$ (where $D' = L, R$), in the background sourced by particle 1 with mass $m_1$, located at coordinate $x_1^{D}$ (where $D = L , R $). Denoting the relative distance between them as $x_0 = | x_2^{D'} - x_1^D |$, this is given by
\begin{align}
    \hat H &= \sqrt{- g_{00}(x_0)  \big( m_2^2 + c^2 g_{ij}(x_0) P^i P^j \big) }
    \nonvt \\
    &\to \sqrt{- g_{00} m_2^2}
    \label{eq31}
\end{align}
where the notation $g_{00} (x_0), g_{ij}(x_0)$ denotes that the metric components are functions of $x_0$, and $P^i$, for $i = 1,2,3$ are the components of the momentum operator (which we set to zero in the second line, since Refs.\ \cite{bose2017spin,marletto2017gravitationally,christodoulou2019possibility} assume fixed trajectories). Now, for the post-Newtonian background introduced earlier, Eq.\ (\ref{eq9}), with components  
\begin{align}
    g_{00} &= - \bigg( 1 + \frac{2 \phi(x_0)}{c^2} \bigg) 
    \vt \\
    g_{ij} &= \delta_{ij} \bigg( 1 - \frac{2 \phi (x_0)}{c^2} \bigg) 
    \vt 
\end{align}
Eq.\ (\ref{eq31}) simplifies to, 
\begin{align}
    \hat H &\simeq mc^2 \bigg( 1 + \frac{\phi (x_0) }{c^2} \bigg) 
    \label{eq34}
\end{align}
where the first term in the brackets will only contribute a global phase to the dynamics.

Before computing the dynamics, let us observe that the states $| x_i^L \rangle$, $| x_i^R\rangle$ are related by a unitary operator $\hat{T}(X_1) \equiv \hat T_1(X_1)  \otimes \hat T_2(X_1)$
(acting jointly on both particles) implementing a translation by the distance $X_1$, namely,
\begin{align}
    | x_1^R, x_2^L \rangle &= \hat T^\dd (X_1) | x_1^L , x_2^L - X_1 \rangle     
    \nonumber \vt 
    \\
    | x_1^R , x_2^R \rangle &=\hat T^\dd (X_1) | x_1^L , x_2^R - X_1 \rangle 
    \nonumber \vt 
\end{align}
so that we can express the state of the particles alone from Eq.~(\ref{eq22initialstate}) as 
\begin{align}
\label{eq16}
    & | \psi_\mrm{GIE} \rangle \equiv  | \bar \psi_\mrm{GIE} \rangle  = \frac{1}{2} ( | x_1^L , x_2^L \rangle + | x_1^L, x_2^R \rangle ) 
    \nonumber \vt \\
    & \:\:\: + \frac{\hat{T}^\dd}{2} ( | x_1^L , x_2^L - X_1 \rangle + | x_1^L , x_2^R - X_1 \rangle )
    \vt 
\end{align}
where $| \bar \psi_\mrm{GIE} \rangle \equiv | \psi_\mrm{GIE} \rangle$ as per our previous nomenclature for equivalent representations of the state, and we have suppressed the coordinate label of $\hat T^\dd$ for brevity.\footnote{We remark that in Eq.\ (\ref{eq16}) one could in-principle include, in keeping with the notation of Eq.\ (\ref{eq14}) and Ref.\ \cite{christodoulou2019possibility}, the metric sourced jointly by both particles (but with the first particle fixed at $x_1^L$):
\begin{align}
    | \bar \psi_\mrm{GIE}^{(g)} \rangle &= \frac{1}{2} ( | g (x_1^L , x_2^L ) , x_1^L , x_2^L \rangle + | g(x_1^L, x_2^R) , x_1^L, x_2^R \rangle 
    \nonvt \\
    & + \frac{\hat T^\dd}{2} ( | g(x_1^L, x_2^L-X_1) , x_1^L, x_2^L - X_1 \rangle 
    \nonvt \\
    & + | g(x_1^L , x_2^R - X_1 ) , x_1^L , x_2^R - X_1 \rangle ) 
    \vphantom{\frac{\hat T^\dd}{2} }
    \label{eq32}
\end{align}where $\hat T^\dd$ acts on both the metric and position DoFs. However we strongly emphasise that no effect unique to a quantum theory of gravity (i.e.\ a quantum superposition of metrics) could be witnessed through an interference experiment involving a state of the form Eq.\ (\ref{eq32}). This is because the GIE setup only involves two particles mutually interacting via gravity, and as we demonstrate in our analysis below, this means that the resulting phase shift (and any observable) can be reproduced in the interpretation that one of the particles (here particle 1, though this choice is arbitrary) sources a classical gravitational field for the other, with suitably modified initial and final states.} Equation (\ref{eq16}) tacitly includes the metric sourced by particle $i = 1$, though since it is on a classical trajectory, one need not invoke a metric Hilbert space in this representation of $| \psi_\mrm{GIE} \rangle$.

Now, in the GIE proposal, after some interval of time-evolution (under which part of the state acquires a relative phase) each particle is measured in a superposition by an uncorrelated measuring device. Following our previously introduced notation, we represent this measurement as  
\begin{align}
    | \Omega_\mathrm{DoF}(x_1,x_2) \rangle \equiv | \Omega(x_1,x_2) \rangle = | \Omega_1(x_1) \rangle | \Omega_2(x_2) \rangle \nonumber 
\end{align}
denoting the two particles.

To proceed, let us compute the time-evolution generated by the Hamiltonian in Eq.\ (\ref{eq34}), which for $\phi(x_0)$ being the Newtonian potential, reduces to Eq.\ (\ref{eq27newtonianinteraction}). The initial state, Eq.\ (\ref{eq16}), becomes, 
\begin{align}\label{eq:NewtonianUnitary}
    & \hat U  | \bar \psi_\mrm{GIE} \rangle = \frac{\hat U}{2} (  | x_1^L , x_2^L \rangle + | x_1^L , x_2^R \rangle ) 
    \nonumber \vt \\
    & \:\:\: + \frac{\hat U \hat T^\dd}{2} ( | x_1^L, x_2^L - X_1 \rangle 
    + | x_1^L, x_2^R - X_1 \rangle ) 
    \vt 
\end{align}
Next, using the general steps outlined in the framework introduced in Sec.\ \ref{sec:IIIA} and \ref{sec:IIIB}, we obtain the relevant probability amplitude by projecting the two-particle state onto $| \Omega \rangle \equiv | \Omega_1 (x_1) \rangle | \Omega_2 (x_2) \rangle \equiv | \Omega_1 \rangle | \Omega_2 \rangle $:
\begin{align}
    & \langle \Omega | \hat{U} | \bar \psi_\mrm{GIE} \rangle 
    = \frac{1}{2} \langle \Omega_1 | \langle \Omega_2 | \hat U ( | x_1^L , x_2^L \rangle 
    +  | x_1^L, x_2^R \rangle )
    \nonumber 
    \vt \\
    & \:\:\: + \frac{1}{2} \langle \hat T \Omega_{1} | \langle \hat T \Omega_{2} | \hat T \hat{U} \hat T^\dd | x_1^L, x_2^L - X_1 \rangle 
    \nonumber \vt  \\
    & \:\:\: + \frac{1}{2} \langle \hat T \Omega_1 | \langle \hat T \Omega_2 | \hat T \hat U \hat T^\dd | x_1^L , x_2^R - X_1 \rangle ) 
    \label{eq32}
    \vt 
\end{align}
where we have inserted the identity in the latter two amplitudes. We have denoted the action of the translation operator on the final state of the measurement as 
\begin{align} 
\langle \hat T \Omega_1 | \langle \hat T \Omega_2 | &\equiv \langle \Omega_1(x_1 - X_1) | \langle \Omega(x_2 - X_1) | 
\nonvt \\
&= \langle \Omega_1 | \langle \Omega_2 | \hat T^\dd 
\nonvt 
\end{align} 
in accordance with the nomenclature of Eq.\ (\ref{eq8nomenclature}) and (\ref{eq8diffeo}), where $x_1, x_2$ generically label the positions of particles $1,2$ (below we consider measurements performed on each particle in a superposition of paths). Note also that since the Newtonian potential is invariant under passive translations, $\hat T \hat U  \hat T^\dd = \hat U $, as the operator $\hat T$ shifts the coordinates of both particles by the same distance. Finally, we emphasize that the effect of $\hat U $ is to introduce a relative phase to the third term of Eq.\ (\ref{eq32}), $\phi = G m_1 m_2t /\hslash d$, which is identical to that obtained in the original formulation of the protocol under the assumptions stated previously \cite{bose2017spin,marletto2017gravitationally}. With this understanding, we leave Eq.\ (\ref{eq32}) with the evolution operators acting on the states in order to finalize our argument. 

Consider now that the measurements are performed in a superposition of the paths, i.e.
\begin{align}
    |\Omega_{i}(x_i) \rangle = \frac{1}{\sqrt{2}} ( | x_i^L \rangle + | x_i^R \rangle ).
    \nonumber 
\end{align}The final probability amplitude in the original formulation i.e.\ Eq.\ (\ref{eq22phaseshift}), reads 
\begin{align}
    \langle \Omega | \hat{U} | \psi_{\mrm{GIE}} \rangle 
    =\frac{1}{4} \sum_{A,B} \langle x_1^A, x_2^B | \hat U | x_1^A, x_2^B \rangle  
    \vt 
    \label{eq27gie}
\end{align}
where the sum $A, B = L , R$ runs over the four relevant amplitudes. We remark that this is Eq.\ (\ref{eq22phaseshift}) with the inclusion of the final measurement; as discussed above, we retain the operatorial form of the evolution operators Eq.\ (\ref{eq27gie}) with the understanding that the third term accumulates a relative phase given in Eq.\ (\ref{eq22phaseshift}). 

Now using our formalism, i.e.\ Eq.\ \eqref{eq32}, which requires no reference to the metric sourced by the particles, and orthogonality of different position eigenstates,  the same amplitude reads 
\begin{equation}\label{eq:BMVamplitdue_classical}
    \langle \Omega | \hat{U}   | \bar \psi_\mathrm{GIE} \rangle = \langle \bar x_1, \bar x_2 | \hat U  | \bar x_1, \bar x_2 \rangle, 
    \vt 
\end{equation}
where we have defined 
\begin{align} 
    & | \bar x_1, \bar{x}_2\rangle := \frac{1}{2} | x_1^L \rangle \otimes ( | x_2^L \rangle + | x_2^L - X_1 \rangle ) 
    \nonvt \\
    & \:\:\:\:\: + \frac{1}{2} | x_1^L \rangle \otimes ( | x_2^R \rangle  + |   x_2^R - X_1 \rangle) .
\vt 
\end{align}
Clearly we can write Eq.\ (\ref{eq:BMVamplitdue_classical}) as $\langle \bar{x}_2 | \hat{U}  ( \bar x_1 ) |\bar{x}_2 \rangle$ where
\begin{align} 
    \hat{U}   (\bar x_1) \equiv \langle \bar x_1 | \hat{U}   | \bar x_1 \rangle = \int \D x_2 \: e^{\frac{i Gm_1m_2 t}{\hslash | x_1^L - \hat{x}_2|}} |x_2\rangle\langle x_2| 
    \vphantom{\Bigg)}
\end{align}
is the gravitational potential sourced by particle $1$ positioned at $x_1^L$ and acting on the states of particle $2$. Note that our nomenclature $| \bar x_1 \rangle \equiv | x_1^L \rangle$ denotes the transformed final state in which particle 1 is measured, as per our formalism.

The calculation above shows that GIE, originally explained as due to a quantum superposition of  gravitational fields, and subsequently as a superposition of two-body spacetime metrics, can be explained in terms of a test mass prepared and measured in a superposition of four paths in the presence of a classical potential sourced by the other mass. Thus, there is an inherent ambiguity in the interpretation of such experiments as testing quantum features of gravitational degrees of freedom. Indeed, the latter interpretation is a variation of the celebrated COW (Collela-Overhauser-Werner) experiment \cite{collelaPhysRevLett.34.1472}, in which neutrons in a Mach-Zehnder interferometer oriented within the Earth's classical gravitational field experienced a gravitationally-induced phase shift dependent on the arm height. We stress that this ambiguity is present because in the considered scenario only relative distances between the particles play a role in determining the final probability amplitude, and not their absolute locations relative to, say, the laboratory reference frame.

Consequently, the interpretation that the initial state Eq.\ \eqref{eq14} evolves into Eq.\ (\ref{eq15}) describing a superposition of semiclassical metrics sourced by the composite system comprised of both particles (the interpretation given by the authors of Ref.\ \cite{christodoulou2019possibility}) while correct, is not required to predict the amplitude of interest in this proposal. Witnessing the effect of the superposition, in particular that described by the state given in Eq.\ (\ref{eq15}) or equivalently Eq.\ (\ref{eq32}), would require a third test system residing in the spacetime jointly sourced by the two particles. In such a scenario, it is impossible to classically fix the positions of two of the systems via quantum coordinate transformations.

Instead, the result of the experiment can be adequately described in terms of a superposition of distances between masses where the one designated as the source is fixed. Consequently, if we wanted to explicitly include a metric sourced by particle $1$, it likewise could be interpreted as fixed. Our analysis emphasizes that any interpretation of a physical process should not be based on a particular mathematical representation of the states--which as highlighted above, is inherently ambiguous for any superposition of amplitudes differing by a symmetry of dynamics.

Removing the above ambiguity requires superpositions of non-diffeomorphic metrics. By definition such metrics are not related by any change of coordinates, since they effectively give rise to unique solutions to the Einstein field equations. It then follows that the prior analysis, in which various amplitudes in superposition could be related by coordinate transformations, is no longer applicable. In such cases, the superposition of geometries that arises is unambiguously quantum-gravitational, insofar as the resulting metrics associated with different amplitudes cannot be re-expressed as a single classical metric with suitably transformed dynamics and measurement bases for the remaining DoFs. Applied to the GIE proposals, this could involve atoms that are prepared in a superposition of energy eigenstates \cite{Zych2011,ruiz2017}, with each eigenstate generating a different curvature (such scenarios would likely be of conceptual, rather than of practical significance). Related examples that have been studied in the literature include a black hole \cite{fooPhysRevLett.129.181301,foodoi:10.1142/S0218271822420160} or dark matter distribution \cite{allaliPhysRevLett.127.031301} in a superposition of masses, and an expanding universe in a superposition of expansion rates \cite{Foo_2021}.

\subsection{Apparent Decoherence of Black Hole Superpositions}\label{sec:apparent}

In this section, we apply our approach to scenarios exploring the phenomenon of decoherence. We show how following a common approach may lead to different conclusions about the decoherence of a mass configuration, depending on how one defines the coordinate system to describe the configuration.
Let us consider for example, the case of a black hole at some position and its Hawking radiation, described by  the state $| \chi_{\mrm{R}} \rangle \equiv | \chi_{\mrm{R}}(\mathbf{x})\rangle$ with $\MB{x}$ being the spatial 3-vector of the black hole \cite{arrasmith2019}. In this picture, a black hole in a superposition of different positions $\MB{x}, \MB{y}$ i.e.\ sourcing the metrics $g(\MB{x}), g(\MB{y})$, becomes entangled with its own radiation  
\begin{align}\label{10}
    | \psi \rangle &= \frac{1}{\sqrt{2}} ( | g(\MB{x}) , \chi_\mrm{R}(\MB{x}) \rangle + | g(\MB{y}) , \chi_{\mrm{R}}(\MB{y}) \rangle 
\end{align}
Upon tracing out the radiation degrees of freedom from the state in Eq.~\eqref{10}, one finds
\begin{align}\label{eq22decoherence}
    \mathrm{Tr}_\mrm{R} | \psi \rangle \langle \psi | &= \frac{1}{2} \begin{pmatrix}
    1 & \nu
    \\
    \nu^\star & 1
    \end{pmatrix} \to \frac{1}{2} \begin{pmatrix}
    1 & 0 \\ 0 & 1 
    \end{pmatrix},
\end{align}
where $\nu = \langle \chi_\mrm{R}(\MB{x}) | \chi_\mrm{R}(\MB{y}) \rangle$ is the overlap between the radiation states associated with the different positions of the black hole. For distinguishable states of the radiation, over time the off-diagonal elements of the reduced density matrix will become suppressed, leaving a classical mixture of the two locations of the black hole. This has been argued 
to lead to fundamental decoherence \cite{arrasmith2019,Demers_1996,Gambini2007} , due to the 
fact that a black hole cannot be isolated from its Hawking radiation and that the state of radiation depends on the location of the black hole.

However, 
the relevant question is, with respect to what   
physical system is the black-hole-radiation system superposed? Clearly, choosing coordinates whose origin is defined as the location of the black hole, often assumed as a  natural choice,  Eq.\ (\ref{10}) by construction becomes a product state of the black hole and its radiation,
\begin{align}\label{eq22}
    |\psi\rangle &= | g(\MB{x}) , \chi_{\mrm{R}}(\MB{x}) \rangle ,
    \vt 
\end{align}
from which one would not deduce decoherence. Based on our framework for the relativity of superpositions, the transformation between Eq.~(\ref{10}) and Eq.~(\ref{eq22}) involves or acts on all DoFs. The tacit assumption behind Eq.~(\ref{eq22decoherence}) is that in computing the overlap $\nu$, one necessarily has access to an uncorrelated matter system (``measuring device(s)''), for example, a first-quantized particle that that can 
interact with the radiation field $\chi_\mrm{R}(\MB{x})$ \cite{wojciechRevModPhys.75.715,schlosshauerRevModPhys.76.1267}.

To see this, it is instructive to consider probability amplitudes that are obtained by coupling a measuring device whose position is correlated with the position of the black hole (the measuring device here is  a proxy for arbitrary matter DoFs). Labelling the states of this device with its position $x$ and some internal state $\phi_\mrm{I}$, i.e.~$|\phi_\mrm{I}(\MB{x}) \rangle$, let us consider the state of this device and the BH--radiation system  
\begin{align}\label{eq24}
    & \frac{1}{\sqrt{2}} | g(\MB{x}) , \chi_{\mrm{R}}(\MB{x}) \rangle | \phi_\mrm{I}(\MB{x} + \MB{d} ) \rangle \nonumber \vt  
    \\
    & \qquad + \frac{1}{\sqrt{2}} | g(\MB{y}) , \chi_{\mrm{R}}(\MB{y}) \rangle | \phi_{\mrm{I}}(\MB{y} + \MB{d}) \rangle ,
    \vt 
\end{align}
where the distance between the device and the black hole, $| \mathbf{d} |$, is identical in each branch of the superposition. Moreover, we assume that the two black hole positions are separated by a distance $| \MB{X} |$ and hence the states are related by the unitary transformation, 
\begin{align}\label{eq:translation:45}
  &  | g(\MB{y}) , \chi_{\mrm{R}}(\MB{y}) \rangle | \phi_\mrm{I}(\MB{y} + \MB{d} ) \rangle \nonumber =
  \vt \\
  & \qquad \hat{T}^\dd(\MB{X}) | g(\MB{x}) , \chi_{\mrm{R}}(\MB{x}) \rangle | \phi_\mrm{I}(\MB{x} + \MB{d} ) \rangle. 
  \vphantom{ g)}
  \vt 
\end{align}Time evolution of the state is given by a unitary of the general form introduced in the earlier sections,
\begin{align} 
    \hat{U}  = \hat U(\MB{x}) | g(\MB{x}) \rangle\langle g(\MB{x}) | + \hat U(\MB{y}) | g(\MB{y}) \rangle\langle g(\MB{y}) | 
    \label{eq33ut}
\end{align}and where each $\hat U(i)$ describes an interaction between the measuring device and the radiation field for the $i$th position of the black hole. Note that here the metric in each amplitude remains stationary. Thus, applying Eq.~\eqref{eq33ut} to evolve the initial state gives
\begin{align} 
    & \frac{1}{\sqrt{2}} \hat U(\MB{x}) | g(\MB{x}), \chi_\mrm{R}(\MB{x}) \rangle | \phi_\mrm{I}(\MB{x}+\MB{d}) \rangle 
    \nonumber \vt 
    \\
    & \qquad + \frac{1}{\sqrt{2}} \hat U(\MB{y}) | g(\MB{y}) , \chi_\mrm{R}(\MB{y}) | \phi_\mrm{I}(\MB{y}+\MB{d})\rangle .
   \vt  
   \label{eq40}
\end{align}
Using Eq.~\eqref{eq:translation:45}
and denoting $|\phi'_\mrm{I}(\MB{x}+\MB{d})\rangle$  the state of  the measuring device placed at distance $| \MB{d} |$ from the black hole after it has interacted with the radiation field, we find
\begin{align}\label{eq26}
    & \frac{1}{\sqrt{2}} | g(\MB{x}) , \chi_{\mrm{R}}(\MB{x}) , \phi_{\mrm{I}}'(\MB{x}+\MB{d}) \rangle 
    \nonumber \vt 
    \\
    & + \frac{1}{\sqrt{2}} \hat{T}^\dd(\MB{X}) | g(\MB{x}) , \chi_{\mrm{R}}(\MB{x}) , \phi_{\mrm{I}}'(\MB{x} + \MB{d} )\rangle .
\end{align}
This state factorizes for a scalar internal DoF such as rest mass-energy, i.e.\ since $| \phi_\mrm{I}(\MB{x}+ \MB{d} ) \rangle = | \phi_\mrm{I}( \MB{y} + \MB{d} ) \rangle$ and $| \phi_\mrm{I}'(\MB{x} + \MB{d}) \rangle = \h T^\dd_\mrm{I}(\MB{X}) | \phi_\mrm{I}' ( \MB{x} + \MB{d} ) \rangle$ are unaffected by translations (nor will the final measurement basis $| \Omega_\mrm{I}(\mathbf{x}+\mathbf{d})\rangle$, see Eq.\ (\ref{eq27}) and (\ref{eq36finalprob})), so the total probability to find the internal DoF in some final state $|\Omega_\mrm{I}(\MB{x}+\MB{d}) \rangle$ reads
\begin{align}\label{eq27}
   & \frac{1}{2} \Big| g(\MB{x}), \chi_{\mrm{R}}(\MB{x}) \rangle + \hat{T}^\dd(\MB{X})|g(\MB{x}) , \chi_\mrm{R}(\MB{x})\rangle \Big|^2 
   \nonumber \vt 
   \\
   & \qquad \times \Big| \langle\Omega_\mrm{I}(\MB{x}+\MB{d})| \phi_\mrm{I}'(\MB{x}+\MB{d})\rangle \Big|^2
   \vt 
\end{align}
Assuming the orthogonality for position states of the black hole implies that 
\begin{align}
    \langle  \chi_\mrm{R}(\MB{x}), g(\MB{x}) | \hat{T}(\mathbf{X}) | g(\MB{x}) , \chi_{\mrm{R}}(\MB{x}) \rangle = 0 
    \vt \nonumber 
\end{align}which further means that Eq.~\eqref{eq27} simplifies to
\begin{align}
     & \Big| \langle \chi_{\mrm{R}}(\MB{x}), g(\MB{x})  | g(\MB{x}) , \chi_{\mrm{R}}(\MB{x}) \rangle \Big|^2 
     \Big | \langle \Omega_{\mrm{I}}(\MB{x}+\MB{d}) | \phi'_{\mrm{I}}(\mathbf{x}+\mathbf{d}) \rangle \Big|^2 
     \nonumber \vt \\
     & \qquad = \Big | \langle \Omega_{\mrm{I}}(\MB{x}+\MB{d}) | \phi'_{\mrm{I}}(\MB{x}+\MB{d}) \rangle \Big|^2 
     \label{eq36finalprob}
\end{align}
which is exactly the same probability as we get in the case when the positions of the measuring device, the black hole, and the radiation field are fixed i.e.~the black hole has a ``classical'' relative distance with respect to the measuring device (note that for a coordinate-dependent internal DoF like spin, Eq.\ (\ref{eq27}) will no longer factorize--in such a case, the measurement can be interpreted as occurring in a correlated manner analogous to Eq.\ (\ref{eq12}). This does not affect the conclusion of our analysis, Eq.\ (\ref{eq36finalprob})). If we consider all other DoFs in the universe are likewise correlated, all observables will be consistent with a classical position of the black hole and will not allow us to conclude that there is any decoherence.

We note that the assumption that the radiation states $|\chi_i\rangle$ are related by a translation is consistent with the treatment in Ref.~\cite{arrasmith2019} where the density matrix of the radiation in the momentum basis is taken to be 
\begin{align} 
    | \chi_{\mrm{R}}(\MB{x}) \rangle\langle \chi_{\mrm{R}} (\MB{y})  | = \int\D^3 \mathbf{k} \frac{p(| \MB{k} | )}{4\pi^2 |\MB{k} |} e^{-i\mathbf{k}\cdot \mathbf{x}} | \mathbf{k} \rangle\langle \mathbf{k} | e^{i\mathbf{k}\cdot \mathbf{y}} ,
    \label{eq49}
\end{align}
where $p(| \MB{k} |)$ is the probability of emitting a particle with momentum $| \mathbf{k} |$ and $\mathbf{x},\mathbf{y}$ are the spatial 3-vectors associated with the coordinate systems $x,y$ respectively.\footnote{Equation (\ref{eq49}), taken from Ref.\ \cite{arrasmith2019}, assumes plane waves for the radiation in the region asymptotically far from the black hole.} This in particular implies that  $\langle \chi_{\mrm{R}} (\MB{y}) |\chi_{\mrm{R}}(\MB{x}) \rangle$ is only a function of $|\mathbf{x}-\mathbf{y}|$, see Ref.\ \cite{arrasmith2019} for an explicit expression. 

The alternative scenario is to consider a measuring device whose position is not  correlated with the position of the black hole, for example described by the state \begin{equation}\label{eq29_main}
\frac{1}{\sqrt{2}}\left( | g(\MB{x}) , \chi_\mrm{R}(\MB{x})\rangle  + | g(\MB{y}) , \chi_\mrm{R}(\MB{y}) \rangle \right) |\phi_\mrm{I}(\MB{z}) \rangle 
\end{equation}
where the position of the device $\MB{z}$ is uncorrelated with the position of the black hole, and so the device's relative distance to the black hole is now different in each amplitude, i.e.~ $ \MB{z}  =  \MB{x} + \MB{d}_1  = \MB{y} + \MB{d}_2 $ with $\MB{d}_1 \neq \MB{d}_2$. In this case, Eq.~\eqref{eq26} reads
\begin{align}
     & \frac{1}{\sqrt{2}}| g(\MB{x}) , \chi_\mrm{R}(\MB{x}) | \phi_\mrm{I}'(\MB{x}+\MB{d}_1) \rangle \nonumber
     \\
     & + \frac{1}{\sqrt{2}} \hat{T}^\dd(X)|g(\MB{x}) , \chi_{\mrm{R}}(\MB{x}) \rangle |\phi_\mrm{I}'(\MB{x}+\MB{d}_2)\rangle,
     \nonumber 
\end{align}
and the state of the measuring device in general gets correlated with the black-hole-radiation system,
which in turn leads to suppression of the off-diagonal terms of the black-hole radiations system given by the overlap between the states of the measurement device $\langle\phi'_\mrm{I}(\mathbf{x}+\mathbf{d}_1))|\phi'_\mrm{I}(\mathbf{x}+\mathbf{d}_2)\rangle$. Here we can conclude that there will be  decoherence over time as information is exchanged between the black hole and the measuring device through the radiation at the location of the device. Note however, that for any translation invariant interaction, the information is only encoded in the distances $\mathbf{d}_i$ between the device and black hole, i.e.~for any $\mathbf{d}$ we have $\langle\phi_\mrm{I}'(\mathbf{d})|\phi_\mrm{I}'(\mathbf{d})\rangle = 1$.

The conclusion is therefore that one should rather speak about the decoherence of the relative distance between a black hole and other systems that can obtain information about the black-hole-radiation system. In this sense, decoherence of black holes due to Hawking radiation does not need to be interpreted as fundamental, since in the absence of any external DoFs, the position of the black hole is not even unambiguously defined. This is related to the point made by Unruh and referred to as ``false loss of coherence'' \cite{unruh2000false}. There, the observation is that decoherence is often inferred by virtue of the coupling between a system and its environment; however as long as any changes to the system are made adiabatically, when the superposed amplitudes are brought together the state of the environment will also be brought together and will not lead to decoherence. Here, however, we do consider that the radiation keeps being emitted over time. The two scenarios are then: is there any external system that can read out the difference as in Eq.~\eqref{eq29_main} or is all external matter also correlated with the black hole as in Eq.~\eqref{eq24}? In the latter case, the superposition state  can be regarded as only due to a choice of coordinates.

\subsection{Coherence of Spacetime Superpositions}\label{secIIe}

We have argued in the preceding section that the presence of matter that can encode information about the black hole is crucial to consider even if the DoFs of the radiation from the black hole are included in the analysis.  We now apply our framework to analyze quantitatively how much spacetime superpositions decohere due to the presence of such matter DoFs.

Let us consider for concreteness the case where a probe particle resides outside a black hole in a superposition of relative distances from the horizon. For simplicity we model the particle as a qubit with scalar-valued internal states $| \phi_\mrm{I}\rangle \equiv | \phi_\mrm{I}(\MB{z}) \rangle \in \{| 0\rangle,  |1 \rangle \}$ (where $\mathbf{z}$ is the 3-position of the probe), whose evolution depends on the metric and the probe's position through interaction with a quantum field $\hat{\phi}$, whose initial state is denoted by $| \phi_\mrm{F} \rangle$. To summarise, the initial state  of the black hole, probe internal DoF, and quantum field is given by
\begin{align} 
    |\psi\rangle=\frac{1}{\sqrt{2}} (| g(\MB{x}) \rangle + | g(\MB{y}) \rangle )  | \phi_\mrm{I} \rangle |\phi_\mrm{F}\rangle
\nonumber 
\end{align}
where in keeping with the nomenclature of Sec.\ \ref{sec:apparent}, $\MB{x},\MB{y}$ denote the spatial 3-vector of the black hole (though we could equally replace these with generic coordinate systems following the notation of Sec.\ \ref{sec2b}; see also Appendix \ref{appc}). Like Eq.\ (\ref{eq29_main}), the position of the probe particle is uncorrelated with the black hole's position. The evolution takes the form 
\begin{align}
 \hat{U}     |\psi\rangle = \frac{1}{\sqrt{2}} ( \hat{U}(\MB{x}) | g(\MB{x}) \rangle + \hat{U}(\MB{y}) | g(\MB{y}) \rangle )  | \phi_\mrm{I} \rangle |\phi_\mrm{F}\rangle. 
    \non
\end{align}
Note in addition to the uncorrelated probe state, we assumed that the field $|\phi_\mrm{F}\rangle$ factorizes from the black-hole position. This is a particular choice inspired by a recent investigation of the (2+1)-dimensional Banados-Zanelli-Teitelboim black hole, where the ground state of the field was taken to be the anti-de Sitter vacuum state satisfying this assumption~\cite{fooPhysRevLett.129.181301}. However the method employed here is not constrained by this choice; i.e.~one could as well assume the field state to be correlated with the black hole. 

To compute the decoherence of the black hole, we can consider an interferometric summarisetup in which the black hole (or the control system correlated with the black hole) undergoes a Mach-Zehnder-type trajectory with a controllable phase $\varphi$ on one of the arms \cite{zych2019bell} and is then measured in a superposition basis (in line with the description provided in Sec.\ \ref{sec:IIIB}); for clarity we compute probability amplitudes for the following final superposition state $| \Omega_\mrm{G} \rangle = ( | g(\MB{x}) \rangle + e^{-i\varphi} | g(\MB{y}) \rangle ) /\sqrt{2}$, giving the following conditional state of the internal and field DoFs, 
\begin{align}
    \langle \Omega_\mrm{G} | \h U | \psi \rangle &= \frac{1}{2} ( \h U (\MB{x} ) + e^{i\varphi} \h U ( \MB{y} ) ) | \phi_\mrm{I} \rangle | \phi_\mrm{F} \rangle 
    \vt 
\end{align}
with density matrix, 
\begin{align}
    \rho_\mrm{IF}(t) &= \frac{1}{2} \h U(\MB{x}) \rho_\mrm{IF} \h U^\dd(\MB{x}) + \frac{1}{2} \h U(\MB{y}) \rho_\mrm{IF} \h U^\dd(\MB{y}) 
    \nonvt \\
    & \qquad  + \frac{e^{-i\varphi}}{2} \h U(\MB{x}) \rho_\mrm{IF} \h U^\dd(\MB{y}) + \mrm{h.c} 
    \label{eq52}
\end{align}
where the Hermitian conjugate applies to the last term in Eq.\ (\ref{eq52}). Tracing out the internal and field DoFs leaves
\begin{align}
    \mathrm{Tr}_{\mrm{IF}} \Big[ \langle \Omega_\mrm{G} | \rho(t) | \Omega_\mrm{G} \rangle \Big] &= \frac{1}{2} ( 1 + \four{v} \cos (\varphi) ),
\end{align}
where $\rho(t)$ is the time-evolved state of the full system. We denote $v = |\langle \phi_\mrm{F} | \langle\phi_\mrm{I}   |  \hat{U}(\MB{x})^\dd \hat{U}(\MB{y}) | \phi_\mrm{I}   \rangle |\phi_\mrm{F} \rangle |$ as the interferometric visbility and $\varphi$ is a relative phase, i.e.~$v$ is the contrast between the maxima and minima of the interference fringes obtained as $\varphi$ is varied. If $v = 0$, the black hole state has fully decohered, whereas if $v = 1$, coherence is maximally retained. Importantly, $v$ is explicitly given by the distinguishability of the states of the probe and field DoFs after the interaction. In general, the visibility will take the form (see Appendix \ref{appc}) 
\begin{align} 
v = 1 + L_{12} - \big( P_1 + P_2 \big) / 2 .  
\vt 
\label{eq58}
\end{align}
Here, $L_{12}$ is a contribution originating as an interference term between the superposed amplitudes, while $P_1$ and $P_2$ are contributions from the individual amplitudes themselves. In Appendix \ref{appc}, we give an explicit form for these terms in a particular example of an interaction known as the Unruh-deWitt model \cite{pozasPhysRevD.92.064042,martinPhysRevD.93.044001,Louko_2008}, which can be taken to approximate the light-matter interaction \cite{pozasPhysRevD.94.064074}. In general, $P_{1,2} > L_{12}$ and thus $v < 1$: some amount of coherence is lost. However in the regime of weak coupling between the qubit, field, and spacetime, the decrease in coherence will be very small, $\sim O(\lambda^2)$ where $\lambda \ll 1$. This example is fully consistent with Unruh's observation concerning rapid or strong interactions leading to decoherence, while weak coupling or adiabatic evolution leads to false loss of coherence~\cite{simidzijaPhysRevD.96.065008,simidzijatransmissionPhysRevD.101.036014,gallockyoshimuraPhysRevD.104.125017,tjoaPhysRevD.105.085011}. Moreover, it aligns with the intuition that decoherence results from transfer of information from the system to the environment \cite{unruh2000false}, where a weak coupling allows for some of the coherence to be retained. 

Of course, as more probe particles are introduced into the system, the decrease in visibility becomes more significant. The visibility for a system of $N$ particles coupled to the black hole is given by 
\begin{align}\label{eq29maintext}
    v &= 1 + \frac{1}{2}\sum_{i,j}^N L_{ij} - \frac{1}{2} \sum_{i}^N P_{i}.
\end{align}
Here, $L_{ij}$ is the interference term between the $i$th and $j$th components of the superposition, while $P_{i}$ are the ``local'' contributions from the $i$th amplitude of the superposition. The factor of $(1/2)$ in front of the first summation is to avoid double counting the $L_{ij} = L_{ji}$ terms. Since $P_i > L_{ij}$ for
all $(i,j)$ in general, Eq.\ \eqref{eq29maintext} can be interpreted that as a macroscopic amount of matter is considered, even weakly interacting with the superposition, the decoherence (loss of visibility) becomes very strong.

\section{Discussion}

In this article we have introduced the notion of ``relativity of spacetime superpositions'' for quantum superpositions of source-mass configurations whose amplitudes differ by a coordinate transformation. Such superposition states have garnered significant interest \four{due to} their direct relevance in low-energy tests of quantum gravity, and for understanding emergent quantum-gravitational phenomena from the ``bottom-up'' (i.e.\ investigations of  quantum-gravitational effects that do not rely upon a formal theory of quantum gravity). Our framework only depends upon the basic  tenets of linearity of quantum theory and the invariance of dynamics under coordinate transformations. We have drawn attention to the fact that the choice of labeling configurations of physics systems, even when it involves the notion of superposition or entanglement and the systems are sources of gravitational field, is not fundamental but rather merely conventional. So long as the involved states are related by a  coordinate transformation their associated semiclassical metrics by definition also map to one another. Consequently any such scenario can be mapped to a scenario in which the metric is fixed and classical, while the states and measurements of remaining DoFs are appropriately modified. While such a resulting scenario in general may not appear natural, our point is that it gives rise to the possibility of making equivalent predictions.  Our approach is complementary to recent work in the field of QRFs, as it does not require considering additioanl degrees of freedom of the reference frames, as the linearity of quantum theory and the representation of symmetries via unitary operators is sufficient for the results.

The relativity of spacetime superpositions has significant implications for both conceptual and practical proposals for witnessing quantum-gravitational effects arising from so-called quantum superpositions of geometries. The main point emerging from our framework is that quantum superpositions of gravitational sources whose states are related by a coordinate transformation (which includes all spatial and even temporal superpositions) are not \textit{unambiguously} quantum-gravitational, insofar as they can be described in terms of preparation and measurement of quantum systems on a classical background. This is particularly important in ongoing discussions and proposals for deriving and observing phenomena not describable within the current paradigms of quantum mechanics and classical general relativity.

A further application of our results is in the discussion on decoherence. Our examples show that it is often tacitly assumed but not explicitly stated that the conclusions require considering that specific measurements can be performed on environmental degrees of freedom. Here we have argued that such measurements are not only required to see decoherence, but only through such measurements can superpositions of mass configurations acquire physical meaning. Without them, and assuming complete isolation from all externally coupled systems, superpositions of spatial states of spacetime are operationally indistinguishable from dynamics occurring on a single classical background. 
This highlights the importance of including all interactions with matter DoFs in the analysis of the dynamics of spacetime superpositions, such as those considered in Ref.\ \cite{arrasmith2019} (there, it is the tacitly assumed probe that couples to the position of the gravitational source mass).
However as we have demonstrated, the inclusion of such matter does not make the superposition more fundamentally ``quantum-gravitational,'' for it is again always possible to re-express the spacetime as a single, fixed background and the matter in a corresponding superposition of configurations.

As a final point, and as alluded to above, the symmetry between representations of scenarios involving superpositions of spatial configurations is broken once one considers superpositions of metrics that cannot be related by a diffeomorphism. An example of such a scenario is a source in a superposition of masses,  as the associated states of the gravitational field represent unique solutions to Einstein's field equations and thus \textit{are} physically distinguishable. This point motivates looking for extensions to the current GIE proposals, namely 
experimental schemes involving superpositions of non-diffeomorphic metrics, and the possibility of witnessing quantum-gravitational effects induced thereby.

\section*{Acknowledgements}
We thank Fabio Costa, Germain Tobar, Carolyn Wood, Kacper Debski, and Natalia Moller for useful discussions. 
J.F. is supported by funding provided by the U.S. Department of Energy, Office of Science, ASCR under Award Number DE-SC0023291. M.Z.\ is supported by Knut and Alice Wallenberg foundation through a Wallenberg Academy Fellowship No. 2021.0119. 

\section{Appendix}

\subsection{Coupling a Quantum Field and Matter DoF to a Spacetime Superposition}

In this section, we apply the relativity of spacetime superpositions to the specific interaction between a quantum field and a first-quantized particle with the background sourced by a mass in superposition of configurations. For simplicity, we assume that the particle is pointlike and thereby couples to the field along a semiclassical trajectory.

Consider as usual the initial state of the mass configuration, field, and the particle (i.e.\ matter) to be
\begin{align} 
    | \psi \rangle = \frac{1}{\sqrt{2}} ( | g(x) \rangle + | g(y) \rangle) | \phi_\mrm{M} \rangle | \phi_\mrm{F} \rangle .
    \vt 
    \label{eq46state}
\end{align}where $x,y$ are arbitrary coordinates describing the configurations $g(x), g(y)$. We assume that the particle can be held static in each amplitude of the superposition without necessarily obtaining which-way information about its position relative to the mass. The total Hilbert space of the relevant systems is given by $\mathcal{H} = \mathcal{H}_\mathrm{S} \otimes \mathcal{H}_\mathrm{M} \otimes \mathcal{H}_\mrm{F}$ where $\mathcal{H}_\mathrm{S}$, $\mathcal{H}_\mathrm{M}$, and $\mathcal{H}_\mathrm{F}$ are respectively associated with the spacetime, matter, and field DoFs. 

Let us consider the following general form of the interaction Hamiltonian, $\hat{H}_\mrm{I}$, which couples all degrees of freedom:
\begin{align}\label{eq4ham}
    \hat{H}_\mathrm{I} &= \hat H(x;x_\mrm{M}) \otimes | g(x) \rangle\langle g(x) | 
    + \hat{H} (y;y_\mrm{M}) \otimes | g(y) \rangle\langle g(y) |
\end{align}
where $x_\mrm{M}, y_\mrm{M}$ denotes the worldline along which the particle interacts with the field, which {in general} can  be considered to be the same or different for the two mass configurations; however in Eq.\ (\ref{eq46state}) we assume the former for simplicity. In sum, the Hamiltonian (\ref{eq4ham}) couples the field and matter at location $x_\mrm{M}, y_\mrm{M}$ for the mass configuration $|g(x)\rangle, | g(y) \rangle$. For distinguishable mass configurations the assigned states $|g(x) \rangle$, $|g(y)\rangle$ are mutually orthogonal, and the time evolution operator in the interaction picture between the initial time $t_i$ and final time $t_f$ is given by
\begin{align}
    \hat{U} &=  \exp \left( - i \int_{t_i}^{t_f}\D t \frac{\D \tau}{\D t} \hat{H}_\mrm{I} \right) ,  
\end{align}
where $\tau$ is the proper time of the particle on the worldline associated with $x_\mrm{M}, y_\mrm{M}$. $\hat{U}$ can be written generally as 
\begin{align}
    \hat{U} &= \hat{U}(x;x_\mrm{M}) \otimes | g(x) \rangle \langle g(x) | 
    + \hat{U}(y;y_\mrm{M}) \otimes | g(y) \rangle \langle g(y)  | , 
\end{align}
where $\hat{U}(z;z_\mrm{M})$
is the unitary operator describing the interaction for a given  $| g(z)\rangle$ mass configuration (serving here as a ``control''), \tcb{with $z = x, y$.}

After the interaction and upon measuring the control in the state $| \Omega_\mrm{G} \rangle = ( | g(x) \rangle + |g(y) \rangle )/\sqrt{2}$, the final state of the remaining DoFs (here, the field and the internal state of the particle) is given by
\begin{align}\label{eq:superpos_paths}
    \langle \Omega_\mrm{G} |  \hat{U} | \psi \rangle &= \frac{1}{2} ( \hat{U}(x) + \hat{U}(y)  ) | \phi_\mrm{M} \rangle | \phi_\mrm{F} \rangle ,
    \vphantom{\frac{1}{\sqrt{2}}}
\end{align}
where we have suppressed the dependence on $x_\mrm{M}, y_\mrm{M}$ for brevity. Note that Eq.\ (\ref{eq:superpos_paths}) is 
a conditional 
state of the field and a particle. Considering now that the position of the probe particle is the same in either amplitude, $x_\mrm{M}=y_\mrm{M} = x_0 $ (as assigned by the agent who describes the mass configurations as $g(x), g(y)$), the scenario can be interpreted as the particle at $x$ interacting with a field quantized on a background in a superposition of geometries $g(x)$, $g(y)$. 

An alternative interpretation is to consider the scenario from the ``quantum coordinates'' associated with the mass, in which there is a single, fixed background where the particle is interacting with the field in a superposition of different locations. As argued in the main text, the coordinates assigned to the joint state of the control states with the field $| g(x) \rangle | \phi_\mrm{F} \rangle$, $| g(y) \rangle | \phi_\mrm{F} \rangle$ can be related using an operator $\hat{T}(x,y)$, Eq.\ \eqref{eq7} and \eqref{eq8transformation}, that connects the coordinates parametrizing the two relative configurations of the mass with respect to the matter DoFs. Assuming for simplicity that $g(x), g(y)$ are related by a translation $X$, so that $\hat T(x,y) = \hat T(X)$, we can write
\begin{align} 
    | g(y) \rangle | \phi_\mrm{M} \rangle | \phi_\mrm{F} \rangle &= \hat{T}^\dd(X) | g(x) \rangle | \phi_\mrm{M} \rangle | \phi_\mrm{F} \rangle 
    \non \VT \\
    &= \hat{T}_\mrm{G}^\dd(X) | g(x) \rangle \otimes \hat T^\dd_\mrm{M}(X) | \phi_\mrm{M} \rangle \otimes \hat{T}^\dd_\mrm{F} (X) | \phi_\mrm{F} \rangle
\non \VT 
\end{align}The $\hat{T}(X)$ operator enacts a translation on the entire system. For clarity, we have explicitly written above that it consists here of $\hat{T}_\mrm{G}(X)$, which acts on the source mass and thus also on the associated metric, and $\hat{T}_\mrm{F}(X)$ which acts on the coordinates of the field operator. That is, 
\begin{align} 
\hat{\phi}(x-X) = \hat{T}_\mrm{F}(X) \hat{\phi}(x) \hat{T}^\dd_\mrm{F}(X)
\nonumber .
\vt 
\end{align} 
On the level of the unitary the transformation thus gives 
\begin{align} 
    \hat{U}(x;y_\mrm{M}-X) = \hat{T}(X) \hat{U}(y;y_\mrm{M}) \hat{T}^\dd(X).
    \VT 
\nonumber 
\end{align}
The state of the system is now 
\begin{align}
    \hat{U} | \bar{\psi} \rangle &= \frac{1}{\sqrt{2}} \hat{U}  ( \hat I + \hat{T}^\dd(X)  ) | g(x) \rangle | \phi_\mrm{M} \rangle  | \phi_\mrm{F} \rangle ,
    \nonumber 
\end{align}
where again, we have used the label $| \bar{\psi } \rangle \equiv | \psi \rangle$ to clarify that though this is an equivalent physical situation to that described by Eq.\ (\ref{eq:superpos_paths}), 
the change of the representation of the state from $|\psi\rangle$ to $|\bar{\psi}\rangle$ gives this situation a different interpretation.
After projecting onto the control, again in the superposition state $| \bar{\Omega}_\mrm{G} \rangle = ( \hat I + \hat{T}^\dd(X)) | g(x) \rangle /\sqrt{2}$, we obtain 
\begin{align}
    & \langle \bar{\Omega}_\mrm{G} | \hat{U} | \bar{\psi} \rangle 
    \non 
    \vphantom{\frac{1}{\sqrt{2}}}
    \\
    & \: \: \: = \frac{1}{2} \langle g(x) |  ( \hat I + \hat{T} (X)  ) \hat{U}  ( \hat I + \hat{T}^\dd (X)  ) | g(x)\rangle  | \phi_\mrm{M} \rangle | \phi_\mrm{F} \rangle
    \nonumber .  
\end{align}
Since  
\begin{align} 
    & \hat{T}(x,y)\hat{U}(y;y_\mrm{M})|g(y)\rangle\langle g(y)|\hat{T}^\dd(x,y) 
    \non \VT 
    \\
    & \:\:\: =\hat{U}(x;y_\mrm{M}-X)|g(x)\rangle\langle g(x)|
\nonumber 
\VT 
\end{align}
and  recognising that 
\begin{align} 
    \langle g(x)| \hat{T}(X) \hat{U} | g(x) \rangle = \langle g(x) | \hat{U} \hat{T}^\dd(X) | g(x) \rangle = 0
    \VT 
    \nonumber 
\end{align}due to the assumption of mutual orthogonality of the states $|g(x,y)\rangle$ this simplifies to 
\begin{align}
    & \langle \bar{\Omega}_\mrm{G} | \hat{U}   | \bar{\psi} \rangle 
    \non 
    \vphantom{\frac{1}{\sqrt{2}}}
    \\
    & \:\:\: = \frac{1}{2}  \langle g(x) |  ( \hat{U}(x;x_\mrm{M}) +  \hat{U}(x;y_\mrm{M}-X) |g(x)\rangle | \phi_\mrm{F} \rangle | \phi_\mrm{M} \rangle. 
    \nonumber 
\end{align}
This amplitude can be interpreted as the field and the qubit interacting  on a single metric. Following the special case considered after  Eq.~\eqref{eq:superpos_paths}, when $x_\mrm{M} = y_\mrm{M} \equiv x_0$, we have now that the particle interacts with the field in a superposition of different locations, $x$ and 
$x - X$. Since by construction  $\langle \bar{\Omega}_\mrm{G} | \hat{U} | \bar{\psi} \rangle \equiv \langle \Omega_\mrm{G} | \hat{U} | \psi \rangle$,  physical observables--which could be  transition amplitudes between some states of the field and matter--are invariant between the two representations. This further highlights the equivalence of the two interpretations of this scenario: as a superposition of different states of the source mass with the particle interacting with a field at $x$ or as a single state of the mass configuration with the matter (particle) in a spatial superposition.

\subsection{General Superposition States}

In this section, we generalise the results shown in Sec.\ \ref{relspacesup} to generic superposition states of the metric: 
\begin{align}
    | \psi \rangle &= \beta | g (x_0) \rangle + \sum_{i \neq 0} \alpha_0 | g(x_i) \rangle ,
\end{align}
where $| \beta |^2 + \sum_i | \alpha_i|^2 = 1$ and $x_i$ parametrise the coordinate system of the $i$th spacetime amplitude and $\langle g (x_i) | g(x_j) \rangle = \delta_{ij}$. Again, we assume that the spacetimes are related by diffeomorphisms allowing for the states $| g(x_i)\rangle$ to be related to a background metric state $| g(x_0) \rangle$, via a unitary $| g(x_i) \rangle = \hat T^\dd(x_i,x_0) | g (x_0) \rangle$. As before, we describe some quantum DoFs in the state $| \phi \rangle \equiv | \phi_\mrm{DoF} \rangle$ interacting with the spacetime superposition through the time-evolution governed by $\hat{U} = \sum_{i\neq0} \hat{U}(x_i) | g(x_i ) \rangle\langle g(x_i)|$, before performing a general measurement of the system in the state $| \Omega\rangle$ (describing all DoFs including the metric):
\begin{align}
    & \langle \Omega | \hat{U} | \psi \rangle = \beta \langle \Omega | \hat{U}(x_0) | g(x_0) \rangle | \phi \rangle 
    \nonumber
    \vphantom{\sum_{i\neq0}} \\
    & \:\:\: + \sum_{i\neq 0} \alpha_i \langle \Omega | \hat{U}(x_i) | g (x_i) \rangle | \phi \rangle .
    \nonumber 
\end{align}
Following the same procedure as before, we find that the dynamics can be expressed in such a way as to occur on a single spacetime metric $| g(x_0)\rangle$ while the other DoFs are in a superposition of states and the measurements  are found via the transformations $\hat{T}$:
\begin{align}
    & \langle \Omega | \hat{U} | \psi \rangle = \beta \langle \Omega | \hat{U}(x_0) | g (x_0)\rangle
    | \phi \rangle
    \nonumber
    \vphantom{\sum_{i\neq0}} 
    \\
    & \:\:\: + \sum_{i\neq0} \alpha_i \langle \hat{T}_i \Omega | \hat T_i \hat{U}(x_i) \hat T_i^\dd | g(x_i) \rangle | \hat T_i \phi \rangle .
\end{align}
where $\hat T_i \hat U(x_i) \hat T_i^\dd = \hat U(x_0)$ and the notation $\hat T_i\equiv \hat T(x_i,x_0)$ is understood. Following the main text, let us also look at the scenario where the metric undergoes a complete evolution through an ``interferometric'' setup (i.e.\ where the state of the control or spacetime is measured in an appropriate basis where interference can be witnessed). For an initial state of the spacetime and matter DoFs (the latter in the state $| \phi \rangle \equiv | \phi_\mrm{DoF} \rangle$) given by
\begin{align}
    | \psi \rangle &= \beta | g(x_0) \rangle | \phi \rangle + \sum_{i\neq 0} \alpha_i | g(x_i) \rangle | \phi \rangle  
\end{align}
one can always express the time-evolved state as some modified dynamics on a classical background metric $g(x_0)$:
\begin{align}
    & \hat{U} | \bar{\psi} \rangle = \hat{U} \bigg( \beta + \sum_{i\neq 0} \alpha_i \hat{T}^\dd \bigg) |g(x_0) \rangle | \phi \rangle  
    \nonumber 
    \\
    &\:\:\: = \hat U(x_0) \beta | g(x_0) \rangle | \phi \rangle + \sum_{i\neq 0} \alpha_i \hat U(x_i)\hat T_i^\dd | g(x_0) \rangle | \hat T_i \phi \rangle 
\end{align}
where in the second line we adopt the nomenclature of Eq.\ (\ref{eq8nomenclature}), and we have used the notation $| \psi \rangle \equiv | \bar{\psi} \rangle$ as usual to distinguish the different representations of the same physical situation. Projecting the spacetime (i.e.\ the control) state in the superposition basis, already expressed using the unitary $\hat{T}$ i.e. 
\begin{align}
    | \bar{\Omega}_\mrm{G} \rangle &=  \beta |g(x_0) \rangle + \sum_{i\neq0} \alpha_i \hat{T}_i^\dd| g(x_0) \rangle  
\end{align}
yields the final amplitude
\begin{align}
    & \langle \bar{\Omega}_{\mrm{G}} | \hat{U} |\bar \psi \rangle 
    \non 
    \vphantom{ g)\sum_{i\neq0}}
    \\
    &
    = \langle g(x_0) |  \bigg( | \beta|^2 \hat{U} + \sum_{i,j\neq 0}\alpha_i^\star \alpha_j \hat{T}_i \hat{U}
    \hat{T}_j^\dd  \bigg) |g(x_0) \rangle | \phi \rangle  ,
    \non 
    \\
    & \equiv  \bigg( | \beta|^2 \hat{U}(x_0) + \sum_{i\neq 0} |\alpha_i|^2 \hat{U}(x_i)  \bigg) | \phi \rangle.
\end{align}
This is of course the same result one would obtain if preparing and measuring the spacetime in a superposition of states that are mutually related to each other via some diffeomorphism encoded within $\hat T$. The diffeomorphism invariance of the two scenarios shows that one can always map such a
superposition of geometries--something that is believed to be beyond the formalism of  quantum theory on a fixed background--onto a single spacetime metric.

\subsection{Unruh-DeWitt Model}\label{appc}
In this section, we apply a specific interaction model to calculate the coherence of a generic spacetime superposition coupled to a quantum field and matter DoF modelled as a qubit. We utilize the Unruh-DeWitt model (UdW) used widely in relativistic quantum information and analogue gravity settings. 

For simplicity, let us consider a superposition of two spacetime states $| g(x) \rangle$, $| g(y) \rangle$, such that the initial (product) state of all DoFs is given by $| \psi \rangle = (| g(x)\rangle + |g(y)\rangle ) | 0 \rangle | \phi \rangle/\sqrt{2}$ where $ | 0 \rangle$ is the ground state of the UdW detector and $| \phi \rangle$, the state of the field, is assumed to be the same for either coordinate parametrization $g(x), g(y)$ (though one could in-principle consider detector or field states entangled with the metric). Note also that we use $x,y$ to parametrize two coordinate systems, in keeping with the notation of Sec.\ \ref{sec2b}, though one could also perform this analysis by treating $x\to \MB{x}, y \to \MB{y}$ as describing the position of a source mass, as considered in Sec.\ \ref{sec:apparent}. We work in the interaction picture, with the interaction between all DoFs described by the Hamiltonian,
\begin{align}
    \hat{H}_\mrm{I} &= \lambda \hat{\sigma}(\tau) \eta(\tau)  ( 
    \hat{\phi}(x) \otimes | g(x) \rangle\langle g(x)| 
    \non 
    \vphantom{\frac{1}{\sqrt{2}}}
    \\
    & \qquad + \hat{\phi}(y) \otimes | g(y) \rangle\langle g(y) |  ) .
\end{align}
Here, $\lambda\ll 1$ is a weak coupling constant and
\begin{align} \hat{\sigma} = | 1 \rangle \langle 0 | e^{iE\tau} + \mathrm{H.c.}
\vt 
\end{align} 
is the $SU(2)$ ladder operator between the internal states of the qubit $\{ | 0 \rangle, | 1 \rangle\}$ with energy gap $E$.
The quantities
$x,y$ are the coordinates of the point where the qubit interacts with the field, and may in principle be different on  different branches of the superposition, and  $\eta(\tau)$ is a ``switching function'' parametrized by the proper time of the qubit (its role is defining when the interaction is switched on--it is often chosen to be a Gaussian or a function with a compact support). Note that we have chosen to factor $\hat{\sigma}(\tau)$, $\eta(\tau)$ from the control, which in turn may imply different coordinate times associated with the same $\tau$ on the two manifolds. One can also describe the scenario from the perspective of a common coordinate time (e.g.\ associated with sufficiently far away clock) in which case the individual proper times may be different depending on the position of the qubit relative to the source  mass. 

The time-evolution operator can be expanded in the Dyson series
\begin{align}
    \hat{U} &= \hat I + \hat{U}^{(1)} + \hat{U}^{(2)} + O(\lambda^3) ,
    \nonumber 
\end{align}
up to second-order in the weak coupling constant $\lambda$, where
\begin{align}
    \hat{U}^{(1)} &= - i \infint\D \tau \: \hat{H}_\mrm{I}  (\tau) ,
    \non 
    \\
    \hat{U}^{(2)} &= - \infint \D \tau \int_{-\infty}^\tau \D \tau' \: \hat{H}_\mrm{I}(\tau) \hat{H}_\mrm{I} (\tau') ,
    \nonumber 
\end{align}
are the first- and second-order terms. Our time-evolution takes the usual form,
\begin{align}
    \hat{U} &= \hat{U}(x) \otimes | g(x) \rangle\langle g(x) | + \hat{U}(y) \otimes | g(y) \rangle\langle g(y) | ,
    \vphantom{\frac{1}{\sqrt{2}}} 
    \nonumber 
\end{align}
where
\begin{align}
    \hat{U}(x) &= \hat I + \hat{U}_1 (x)  + \hat{U}_2 (x)  + O(\lambda^3) ,
    \vt 
    \non 
    \\
    \hat{U}(y) &= \hat I + \hat{U}_1(y) + \hat{U}_2 (y) + O(\lambda^3),
    \vt 
    \non 
\end{align}
describe the time-evolutions for the spacetime states $|g(x)\rangle$ and $| g(y) \rangle$ respectively. The time-evolved state is 
\begin{align}
    \hat{\rho} &= \hat{U}   | \psi \rangle\langle \psi | \hat{U}^\dd   
    \vphantom{\frac{1}{2}}
    \non
    \\
    &= \frac{1}{2} \sum_{z,z'} \hat U(z) | g(z) \rangle\langle g (z') | \otimes \rho_\mrm{DF} \hat U^\dd(z') 
    \vphantom{\frac{1}{2}} 
\end{align}
where the sum is over the relevant amplitudes $z,z' = x,y$ and $\hat{\rho}_\mathrm{DF}$ is the density matrix of the detector-field subsystem. The conditional state of the detector-field subsystem, after projecting onto the spacetime state $| \Omega_\mrm{G} \rangle = ( | g(x) \rangle + e^{i\varphi} | g(y) \rangle )/\sqrt{2}$ where $\varphi$ is some controllable phase, is, 
\begin{align}
    \rho_\mrm{DF}(t) &= \frac{1}{2} \hat U(x) \rho_\mrm{DF} \hat U^\dd (x) + \frac{1}{2} \h U(y) \rho_\mrm{DF} \h U^\dd(y) 
    \nonvt \\
    & \qquad + \frac{e^{-i\varphi}}{2} \h U(x) \rho_\mrm{DF} \h U^\dd(y) + \mrm{h.c} 
\end{align}
where $\mrm{h.c}$ denotes the Hermitian conjugate applied to the last term. Tracing over the remaining DoFs gives, 
\begin{align}
    & \mrm{Tr}_\mrm{DF} \Big[ \langle \Omega_\mrm{G} | \rho(t) | \Omega_\mrm{G} \rangle \Big] = \frac{1}{4} \Big( \langle \h U^\dd(x) \h U(x) \rangle + \langle \h U^\dd(y) \h U(y) \rangle 
    \nonvt \\
    & \qquad + e^{-i\varphi} \langle \h U^\dd(y) \h U(x) \rangle + \mrm{h.c} \Big) 
\end{align}
where the expectation values are taken with respect to the detector-field state $| 0 \rangle | \phi \rangle$. Inserting the perturbative expansions for $\h U(x), \h U(y)$ and noting that in general $\langle \h U_2(z) \rangle = \langle \h U_2^\dd(z) \rangle = 2 \langle \h U_1^\dd (z) \h U_1(z) \rangle$ for $z = x,y$, we obtain 
\begin{align}
    & \mrm{Tr}_\mrm{DF} \Big[ \langle \Omega_{\mrm{G}} | \hat{\rho}(t) | \Omega_{\mrm{G}} \rangle  \Big] 
    \non 
    \\
    & \:\:\: = \frac{1}{2} \Big( 1 + \Big( 1 + L_{12} - \big( P_1 + P_2 \big)/2 \Big) \cos( \varphi) \Big) . 
\end{align}
The visibility $v$ is the magnitude of the $\cos(\varphi)$ term, namely
\begin{align}
    v &= 1 + L_{12} -  \left( P_{1} + P_{2} \right) /2
\end{align}
which is Eq.\ (\ref{eq58}) of the main text. The integral forms of $P_i$ and $L_{ij}$ are given by 
\begin{align}
    P_i &= \lambda^2 \int_{-\infty}^{+\infty}\D \tau \D \tau' \chi(\tau) \bar{\chi}(\tau') W(x_i(\tau), x_i'(\tau')) ,
\\
    L_{ij} &= \lambda^2 \int_{-\infty}^{+\infty} \D \tau \D \tau' \: \chi(\tau) \bar{\chi}(\tau') W(x_i(\tau), x_j(\tau')) ,
\end{align}
where $i ,j =1 ,2$, $\chi(\tau) = \eta(\tau) \exp(-iE\tau)$, and the functions $W(x_i,x_j')$ are two-point field correlation functions,
\begin{align}
\label{eq61}
    W(x_i(\tau), x_j(\tau')) &= \langle \phi | \hat{\phi}(x_i(\tau_i))  \hat{\phi}(x_j(\tau_j')) | \phi \rangle .
    \vphantom{\frac{1}{\sqrt{2}}}
\end{align}
where $| \phi \rangle \equiv | \phi_\mrm{F} \rangle$ is the state of the field. The field operators in Eq.\ (\ref{eq61}) are pulled back to the worldline of the qubit as parametrised by the coordinates $x_i$, $x_j'$, which in the case $i \neq j$, describe the different spacetime metrics in superposition, $| g(x) \rangle$, $| g(y) \rangle$. This calculation straightforwardly generalises to additional detectors, giving a visibility of the form Eq.\ (\ref{eq29maintext}). 
\\

\subsection{Analysis of Setup in Ref.\ \cite{belenchiaPhysRevD.98.126009}}

Ref.\ \cite{belenchiaPhysRevD.98.126009} considers how quantizing local vacuum fluctuations (either electromagnetic or gravitational) resolves an apparent causality/complementarity paradox. The setup is as follows: an observer Alice prepares a particle in a spatial superposition with separation $X$, while Bob has control over a particle at some large distance away, that is initially localized by a trap. At the start of the protocol Bob can decide whether or not to release his particle from the trap, while Alice starts to recombine the paths of her particle. An apparent paradox arises when Alice attempts to measure the coherence of the state of her particle, or Bob seeks to gain which-way information about Alice's particle (for full details on the causality/complementarity paradox, see Ref.\ \cite{belenchiaPhysRevD.98.126009}). 

We show below using our framework that whether or not these fluctuations are quantized is independent of the question of whether the metric sourced by Alice's particle is classical or in quantum superposition. Following the argument in Ref. \cite{belenchiaPhysRevD.98.126009}, we first assume that Alice's particle $A$ sources the metric field felt by Bob's particle $B$, in superposition i.e.\ $( | x_\mathrm{A} \rangle | g(x) \rangle + | x_\mrm{A} + X \rangle | g(x + X ) \rangle ) /\sqrt{2}$. Here $| x_\mrm{A} \rangle, | x_\mrm{A} + X \rangle$ are position eigenstates of Alice's particle at $x, x+X$ respectively, and the states $| g(x) \rangle, | g(x+X)\rangle$ live in the metric Hilbert space. The joint state of both particles and the metric sourced by $A$ is (see Fig.\ 1 of Ref.\ \cite{belenchiaPhysRevD.98.126009})
\begin{align}
    | \psi \rangle &=  \frac{1}{\sqrt{2}}( | x_\mrm{A}  \rangle | g(x) \rangle + | x_\mrm{A} + X \rangle | g(x + X ) \rangle ) | f_\mrm{B}(y) \rangle 
    \nonvt \\
    &= \frac{1}{\sqrt{2}} | x_\mrm{A} \rangle | g(x) \rangle | f_\mrm{B} (y) \rangle + \frac{1}{\sqrt{2}} | x_\mrm{A}  + X \rangle | g(x+ X) \rangle  | f_\mrm{B}(y) \rangle 
    \nonvt 
\end{align} 
where $| f_\mrm{B}(y) \rangle$ is the state of Bob's particle (e.g.~$f_\mrm{B}(y)$ can be a Gaussian wavepacket). Time-evolving the state under a quantum-controlled 
unitary 
\begin{align} 
    \hat{U} = \hat{U}(x)\otimes |x_\mrm{A} \rangle\langle x_\mrm{A} | + \hat{U}(x+X) \otimes | x_\mrm{A} +X \rangle \langle x_\mrm{A} + X | 
    \nonvt 
\end{align} 
gives
\begin{align} 
    \hat{U} | \psi \rangle &= \frac{1}{\sqrt{2}} \hat{U}(x) |x_\mrm{A} \rangle | g(x) \rangle | f_\mrm{B}(y) \rangle
    \nonvt 
    \\
    &  + \frac{1}{\sqrt{2}} \hat{U}(x+X) |x_\mrm{A} +X \rangle | g(x+X) \rangle | f_\mrm{B}(y) \rangle 
    \nonvt 
    \\
    &= \frac{1}{\sqrt{2}} \hat{U}(x) | x_\mrm{A}  \rangle  | g(x) \rangle  | f_\mrm{B}(y) \rangle
    \nonvt 
    \\
    & + \frac{1}{\sqrt{2}} \hat{U}(x+X) \hat{T}^\dd(X) | x_\mrm{A} \rangle  | g(x) \rangle| f_\mrm{B}(y-X ) \rangle
    \nonumber 
\end{align} 
where in the second line we utilize the fact that $|x,g(x)\rangle, |x+X,g(x+X)\rangle$ are related by the unitary translation $\hat T(X)$. The particles are then measured in the position basis $ |x_\mrm{A} ' \rangle | y_\mrm{B}' \rangle$:
\begin{align} 
    & \langle x_\mrm{A}' |  \langle y_\mrm{B} ' | \hat{U} | \psi \rangle 
    \nonumber \\ 
    & \:\:\: = \frac{1}{\sqrt{2}} \langle x_\mrm{A}' |  \langle y_\mrm{B}' |  \hat{U}(x) |x_\mrm{A} \rangle | g(x) \rangle | f_\mrm{B}(y) \rangle 
    \nonvt 
    \\
    & \:\:\: + \frac{1}{\sqrt{2}}\langle x_\mrm{A}' |  \langle y_\mrm{B}' | \hat{U}(x+X) \hat{T}^\dd(X) | x_\mrm{A} \rangle | g(x) \rangle  | f_\mrm{B}(y-X) \rangle
    \nonvt 
    \\
    & \:\:\: = \frac{1}{\sqrt{2}} \langle x_\mrm{A}' |  \langle y_\mrm{B}' | \hat{U}(x) |x_\mrm{A} \rangle | g(x) \rangle | f_\mrm{B}(y) \rangle
    \nonvt 
    \\
    & \:\:\: + \frac{1}{\sqrt{2}} \langle x_\mrm{A}'-X |  \langle y_\mrm{B} '-X|  \hat{U}(x)  |x_\mrm{A} \rangle | g(x) \rangle | f_\mrm{B} (y-X) \rangle
    \nonvt 
\end{align}
where to obtain the last line we inserted the identity and used the fact that $\hat{T}(X) \hat{U}(x+X) \hat{T}^\dd(X) = \hat{U}(x)$. The last line is an expression of the fact that the setup in Ref.\ \cite{belenchiaPhysRevD.98.126009} can be understood as occurring on a fixed classical metric described by $| g(x) \rangle$ (which, notably, factors from the above amplitude) while the measurements on the particles $A$ and $B$ occurs in superposition. This in particular entails that the position of the ``screen'' used to measure the position of particle $A$ is, in this form of the amplitude, itself in superposition. The key observation from the above, is that the same discussion of causality based on ``after what time particle $B$ can entangle with $A$ depending on their distance'' applies here, as we have merely re-described the same scenario. Indeed, we have the exact same distances between the particles as in the original formulation and the same probability amplitudes. The difference is that in this picture the positions of the particles are not described by independent measurements but appear correlated. Our analysis above for the gravitational case follows analogously to the electromagnetic case, with a replacement of the gravitational field/metric state with the state of the electromagnetic field.

\vspace{10pt}

\subsection{Dark Matter Scattering}

In this section, we highlight the utility of our framework by applying it to another case of decoherence, namely of general relativistic sources of gravity (such as a dark matter clump) by scattering particles. This scenario has been considered recently in the context of dark matter detection \cite{allaliPhysRevLett.127.031301,Allali_2020}. The authors of \cite{allaliPhysRevLett.127.031301,Allali_2020} consider, in analogy with Eq.\ (\ref{10}), a dark matter (DM) ``Schr\"odinger-cat state'' that interacts with its environment $| \chi \rangle$, leading to an entangled state of the DM and environment:
\begin{align}\label{eq29}
    | \psi \rangle &= \frac{1}{\sqrt{2}} \left( | \mathrm{DM}_1 \rangle | \chi_1 \rangle + | \mathrm{DM}_2 \rangle | \chi_2 \rangle \right). 
\end{align}
The superposition of DM configurations is tacitly defined with respect to some coordinate system in which relative distances between the DM and some incoming particle are expressed. We have demonstrated that the choice of coordinates can change the interpretation of the scenario--here it would be whether decoherence appears in the position of the DM or in the position of the scattered particles. We emphasise that decoherence can be attributed to either of the subsystems involved in the scattering process, implying that it cannot be fundamentally associated with either. Instead, the only physically relevant DoF that decoheres in such a scenario is the relative distance between (in this case) the dark matter and the scattering particles.

The rate of decoherence can be computed through the overlap of the scattering particle states, $\langle \psi_1 | \psi_2 \rangle$ after interaction with the DM states, $|\mathrm{DM}_1 \rangle$, $|\mathrm{DM}_2\rangle$. The computation of the overlap will in general, depend on the convolution of the wavefunctions of the scattered states, given by \cite{allaliPhysRevLett.127.031301},
\begin{align}\label{eq30s12}
    S_{12} &= \int \D^2 b \int \D^3 x \: \psi_{s1}^\star (t,x) \psi_{s2}(t,x) ,
\end{align}
where $b$ is the impact parameter of the scattering process, while $\psi_{s1,2}$ denotes the scattered part of the respective wavefunctions. Importantly, Eq.\ (\ref{eq30s12}) only depends on the wavepacket parameters and the relative distance between the two branches of the DM superposition, which we denote $X_{12}$. Taking into consideration the DM distribution and the spacetime generated by it, and the wavefunction of the scattered particle, this means that such a scenario is physically equivalent to that of a single ``fixed'' DM source from which a particle, in a superposition of relative impact parameters is scattered. The decoherence in this case is not fundamental to either the scattering particle or the DM; both are equivalent situations that depend only on one's choice of coordinates.

\bibliography{main}

\end{document}